\documentclass[twocolumn]{aastex63}

\let\tablenum\relax
\usepackage{siunitx}
\usepackage[version=4]{mhchem}

\usepackage{booktabs}
\usepackage{makecell}
\usepackage{multirow}

\usepackage{epigraph}
\received{24 September 2021}
\revised{30 March 2022}
\accepted{25 April 2022}
\submitjournal{PSJ}

\shorttitle{THAI Part II}
\shortauthors{Sergeev et al.}
\graphicspath{{./}{figures_part2/}}

\begin{document}

\title{The TRAPPIST-1 Habitable Atmosphere Intercomparison (THAI). \\
Part II: Moist Cases -- The Two Waterworlds}

\correspondingauthor{Denis E. Sergeev}
\email{d.sergeev@exeter.ac.uk}

\author[0000-0001-8832-5288]{Denis E. Sergeev}
\affiliation{Department of Mathematics, College of Engineering, Mathematics, and Physical Sciences, University of Exeter, Exeter, EX4 4QF, UK}

\author[0000-0002-5967-9631]{Thomas J. Fauchez}
\affiliation{NASA Goddard Space Flight Center, 8800 Greenbelt Road, Greenbelt, MD 20771, USA}
\affiliation{Goddard Earth Sciences Technology and Research (GESTAR), Universities Space Research Association (USRA), Columbia, MD 7178, USA}
\affiliation{NASA GSFC Sellers Exoplanet Environments Collaboration}

\author[0000-0003-2260-9856]{Martin Turbet}
\affiliation{Observatoire  Astronomique  de  l’Universit\'e  de  Gen\`eve,  Universit\'e  de  Gen\`eve,  Chemin  des Maillettes 51, 1290 Versoix, Switzerland.}
\affiliation{Laboratoire de M\'et\'eorologie Dynamique/IPSL, CNRS, Sorbonne Universit\'e, \'Ecole Normale Sup\'erieure, PSL Research University, \'Ecole Polytechnique, 75005 Paris, France}

\author[0000-0002-1485-4475]{Ian A. Boutle}
\affiliation{Met Office, FitzRoy Road, Exeter, EX1 3PB, UK}
\affiliation{Department of Astrophysics, College of Engineering, Mathematics, and Physical Sciences, University of Exeter, Exeter, EX4 4QL, UK}

\author[0000-0001-5328-819X]{Kostas Tsigaridis}
\affiliation{Center for Climate Systems Research, Columbia University, New York, NY, USA}
\affiliation{NASA Goddard Institute for Space Studies, 2880 Broadway, New York, NY 10025, USA}

\author[0000-0003-3728-0475]{Michael J. Way}
\affiliation{NASA Goddard Institute for Space Studies, 2880 Broadway, New York, NY 10025, USA}
\affiliation{NASA GSFC Sellers Exoplanet Environments Collaboration}
\affiliation{Theoretical Astrophysics, Department of Physics and Astronomy, Uppsala University, Uppsala, Sweden}

\author[0000-0002-7188-1648]{Eric T. Wolf}
\affiliation{Laboratory for Atmospheric and Space Physics, University of Colorado Boulder, Boulder, CO, USA}
\affiliation{NASA NExSS Virtual Planetary Laboratory, Seattle, WA, 98195, USA}
\affiliation{NASA GSFC Sellers Exoplanet Environments Collaboration}

\author[0000-0003-0354-9325]{Shawn D. Domagal-Goldman}
\affiliation{NASA Goddard Space Flight Center, 8800 Greenbelt Road, Greenbelt, MD 20771, USA}
\affiliation{NASA GSFC Sellers Exoplanet Environments Collaboration}
\affiliation{NASA NExSS Virtual Planetary Laboratory, Seattle, WA, 98195, USA}

\author[0000-0002-3262-4366]{Fran\c cois Forget}
\affiliation{Laboratoire de M\'et\'eorologie Dynamique/IPSL, CNRS, Sorbonne Universit\'e, \'Ecole Normale Sup\'erieure, PSL Research University, \'Ecole Polytechnique, 75005 Paris, France}

\author[0000-0003-4346-2611]{Jacob Haqq-Misra}
\affiliation{NASA NExSS Virtual Planetary Laboratory, Seattle, WA, 98195, USA}
\affiliation{Blue Marble Space Institute of Science, Seattle, WA, USA}

\author[0000-0002-5893-2471]{Ravi K. Kopparapu}
\affiliation{NASA Goddard Space Flight Center, 8800 Greenbelt Road, Greenbelt, MD 20771, USA}
\affiliation{NASA GSFC Sellers Exoplanet Environments Collaboration}
\affiliation{NASA NExSS Virtual Planetary Laboratory, Seattle, WA, 98195, USA}

\author[0000-0002-4664-1327]{F. Hugo Lambert}
\affiliation{Department of Mathematics, College of Engineering, Mathematics, and Physical Sciences, University of Exeter, Exeter, EX4 4QF, UK}

\author[0000-0003-4402-6811]{James Manners}
\affiliation{Met Office, FitzRoy Road, Exeter, EX1 3PB, UK}

\author[0000-0001-6707-4563]{Nathan J. Mayne}
\affiliation{Department of Astrophysics, College of Engineering, Mathematics, and Physical Sciences, University of Exeter, Exeter, EX4 4QL, UK}

\begin{abstract}
To identify promising exoplanets for atmospheric characterization and to make the best use of observational data, a thorough understanding of their atmospheres is needed.
3D general circulation models (GCMs) are one of the most comprehensive tools available for this task and will be used to interpret observations of temperate rocky exoplanets.
Due to parameterization choices made in GCMs, they can produce different results, even for the same planet.
Employing four widely-used exoplanetary GCMs --- ExoCAM, LMD-G, ROCKE-3D and the UM --- we continue the TRAPPIST-1 Habitable Atmosphere Intercomparison by modeling aquaplanet climates of TRAPPIST-1e with a moist atmosphere dominated by either nitrogen or carbon dioxide.
Although the GCMs disagree on the details of the simulated regimes, they all predict a temperate climate with neither of the two cases pushed out of the habitable state.
Nevertheless, the inter-model spread in the global mean surface temperature is non-negligible: \SI{14}{\K} and \SI{24}{\K} in the nitrogen and carbon dioxide dominated case, respectively.
We find substantial inter-model differences in moist variables, with the smallest amount of clouds in LMD-Generic and the largest in ROCKE-3D.
ExoCAM predicts the warmest climate for both cases and thus has the highest water vapor content and the largest amount and variability of cloud condensate.
The UM tends to produce colder conditions, especially in the nitrogen-dominated case due to a strong negative cloud radiative effect on the day side of TRAPPIST-1e.
Our study highlights various biases of GCMs and emphasizes the importance of not relying solely on one model to understand exoplanet climates.
\end{abstract}


\section{Introduction} \label{sec:intro}
\epigraph{You must understand, young Hobbit, it takes a long time to say anything in Old Entish. And we never say anything unless it is worth taking a long time to say.}{---J.R.R. Tolkien, \textit{The Two Towers}}

With the advent of powerful new telescopes, such as the James Webb Space Telescope (JWST) and the Extremely Large Telescope (ELT), the characterization of a temperate rocky planet orbiting an M-dwarf star becomes highly likely \citep{Snellen:2015,anglada2016,Lovis:2017,Gillon2017,gillon2020trappist1,Fauchez:2019,Turbet20b}.
In order to make the most use of the vast amount of data these telescopes will collect, a solid understanding of extraterrestrial atmospheres is needed.
Three-dimensional general circulation models (or global climate models, GCMs) represent the complexity of planetary atmospheric dynamics to the best of our present knowledge.
Being complex pieces of code, different GCMs typically produce different results even for the well-constrained climate of Earth \citep[e.g.][]{Webb15,Eyring16,Ceppi17}.
Almost all previous exoplanetary 3D GCM studies have employed only a single GCM due to the challenges of simulating exotic atmospheres on one hand and the exciting variety of possible targets on the other.
It is therefore unclear how accurate predictions from a single GCM will be when used for interpreting observations of a particular exoplanet.

A concerted effort is required to compare predictions of exoplanet GCMs against each other.
Such model intercomparison will provide a more robust reference for explaining the observational data and will serve as a rigorous benchmark for future GCM studies.
Along with theory and abundant observations, model intercomparisons are the bedrock of our understanding of modern and future Earth climate \citep[e.g.][]{Eyring16}.
For exoplanets, observations are extremely scarce, making a model intercomparison an especially valuable tool for our scientific understanding of planetary atmospheres.
The only 3D model intercomparison conducted for a generic exoplanet setup was carried out by \citet{Yang2019}, who reported significant inter-model differences for slowly rotating planets irradiated by M-dwarf stars.

\citet{Yang2019} found that the water vapor radiation feedback was the major cause of the inter-model discrepancy, in agreement with previous 1D intercomparison \citep{yang_differences_2016}; radiative properties of clouds in GCMs were also found important.
More recently, participants in the 2020 THAI workshop noted that inter-model differences in sub-grid parameterizations continue to be a key source of disagreement between simulated climates \citep{Fauchez2021_THAI_workshop}.
The inter-model disagreement due to the treatment of clouds is well-known in Earth climate science.
For paleoclimate reconstructions, it continues to fuel the debate around the Faint Young Sun paradox \citep[e.g.][]{Goldblatt21} and Snowball deglaciations \citep[e.g.][]{Abbot12}.
For future climate projections, it is thought to be the largest source of uncertainty in equilibrium climate sensitivity\footnote{The global mean surface temperature increase after the climate system fully adjusts to a sustained doubling of \ce{CO2} relative to pre-industiral conditions.} \citep{Ceppi17}.
Interestingly, the equilibrium climate sensitivity did not reduce in the latest climate model intercomparison --- Coupled Model Intercomparison Project phase 6 (CMIP6) --- mainly because of stronger positive cloud feedbacks from decreasing low cloud amount and, as a result, albedo of the extratropics \citep{Zelinka20}.
Clouds are tightly coupled with moist convection and so can be affected by convective parameterizations in GCMs \citep{Sherwood14}, though the direct effect of convection on Earth is thought to be of secondary importance \citep{Webb15}.
However, the direct effect of convection on exoplanetary climate predictions from GCMs is still unclear \citep{Sergeev2020}.

The purpose of this study is to analyze inter-model differences for two scenarios of an aquaplanet climate of TRAPPIST-1e.  
In particular, we aim to revisit the moist climate on a tidally locked exoplanet residing within the habitable zone of TRAPPIST-1, an ultracool M-dwarf star, and discuss how moist physics, parameterized differently in different GCMs, affect the temperature regime of the planet, its atmospheric circulation, and cloud variability.
Our results highlight various biases of the THAI GCMs, providing guidance for terrestrial exoclimatology studies, the majority of which so far are done with these four models.
This part of THAI is a logical next step after the dry cases \citep[Part I, see][]{Turbet21_THAI}, increasing the model complexity by including the effects of water vapor and clouds.
Moreover, understanding synthetic observations \citep[Part III, see][]{Fauchez21_THAI} cannot be complete without a thorough understanding of the simulated climates.

This manuscript is structured as follows.
In Sec.~\ref{sec:methods}, we describe the experimental setup and key differences in the representation of moist physics in the four THAI GCMs.
In Sec.~\ref{sec:results} we present the details of the global climate in each group of simulations: \ce{N2}-dominated (Sec.~\ref{sec:hab1}) and \ce{CO2}-dominated (Sec.~\ref{sec:hab2}), starting with the radiative fluxes and thermodynamic profiles and continuing to the cloud cover and atmospheric circulation, before briefly discussing the time variability.
In Sec.~\ref{sec:synthesis}, we discuss overall commonalities and discrepancies between the GCMs.
Sec.~\ref{sec:conclusions} concludes the paper.

\section{Methodology} \label{sec:methods}
\subsection{Planetary configuration} \label{sec:setup}
In this Part II of the THAI trilogy, we conduct moist atmosphere aquaplanet simulations, referred to as Hab~1 \& Hab~2 here and in the THAI protocol \citep{Fauchez2020THAI}.
These simulations are configured to represent two potentially habitable states of TRAPPIST-1e, assuming the planet is tidally locked to its host star (Table~\ref{tab:planet}).
Several modelling studies involving GCMs \citep{Wolf2017,Turbet:2018aa,Fauchez:2019} show that it was possible for TRAPPIST-1e to retain water in its atmosphere or on its surface once it achieved temperate conditions in the habitable zone \citep[HZ,][]{Kopparapu2013}.
Hab~1 is a colder state with a \ce{N2}-dominated atmosphere, 400 ppm of \ce{CO2}, while Hab~2 is a warmer state with a \ce{CO2}-dominated atmosphere.
In both cases, the \ce{H2O} is a condensible species and the mean atmospheric pressure at the surface is 1 bar.
\ce{CO2} condensation has been disabled as not all the four GCMs possess this parameterization.
Hab~1 represents a habitable atmosphere broadly similar to that of modern Earth.
Hab~2 could be representative of the early Hadean period on Earth or Hesperian period on Mars \citep[e.g.][]{Wordsworth16}.
The planet's surface in both Hab~1 and Hab~2 is prescribed to be a slab ocean with the heat capacity of water (\SI{\approx 4e6}{\joule\per\m\cubed\per\K}) and a constant bolometric albedo, 0.06 and 0.25, above and below freezing, respectively.
The slab ocean setup in our GCMs is slightly different, though it does not affect the time averaged steady state of the atmosphere.
ExoCAM and the UM use a single-layer slab ocean, while in LMD-G and ROCKE-3D the slab ocean is split into layers and therefore depends not only on the total heat capacity but also on the thermal inertia of the layers.
The slab ocean depths are as follows: 100 m in ExoCAM, 20 m in LMD-G, 90 m in ROCKE-3D and 1 m in the UM (Table~\ref{tab:grids}).
See \citet{Fauchez2020THAI} for more details on the experimental setup and its motivation.

Both sets of simulations are started from an isothermal (\SI{300}{\K}) dry atmosphere at rest.
The models are integrated until they reach a steady state\footnote{Whether a GCM has reached a steady state is determined by the absence of a long-term trend in variables such as the global mean surface temperature and the top of the atmosphere net energy flux.} and then for a further 100 orbits, with the analysis for this study performed on the data from these 100 orbits (610 Earth days). 
The output is provided every 6 hours in order to sufficiently resolve atmospheric variability.

\begin{deluxetable*}{lll}
\tablecaption{TRAPPIST-1 stellar spectrum and planetary parameters of TRAPPIST-1e \citep{Grimm18}.\label{tab:planet}}
\tablehead{
\colhead{Parameter} & \colhead{Units} & \colhead{Value}
}
\startdata
Star and spectrum &                                & \SI{2600}{\K} BT-Settl with Fe/H=0 \\
Semi-major axis   & AU                             & 0.02928 \\
Orbital period    & Earth day                      & 6.1 \\
Rotation period   & Earth day                      & 6.1 \\
Obliquity         &                                & 0 \\
Eccentricity      &                                & 0 \\
Instellation      & \si{\watt\per\square\meter}    & 900.0 \\
Planet radius     & \si{\km}                       & 5797 \\
Gravity           & \si{\meter\per\second\squared} & 9.12 \\
\enddata
\end{deluxetable*}

\subsection{Models}
As specified in the THAI protocol \citep{Fauchez2020THAI}, four GCMs took part in this study:
\begin{itemize}
    \item the Exoplanet Community Atmospheric Model (ExoCAM, a branch of the National Center for Atmospheric Research Community Earth System Model version 1.2.1),
    \item the Laboratoire de M\'et\'eorologie Dynamique - Generic model (LMD-G),
    \item the Resolving Orbital and Climate Keys of Earth and Extraterrestrial Environments with Dynamics (ROCKE-3D, version Planet\_1.0),
    \item the Met Office Unified Model (UM, science version GA7.0, code version 11.6).
\end{itemize}

\begin{deluxetable*}{lccc}
\tablecaption{Grid resolution and slab ocean setup in THAI GCMs.\label{tab:grids}}
\tabletypesize{\scriptsize}
\tablehead{
\colhead{GCM} & \colhead{\makecell{Number of grid points in the horizontal\\(longitude~$\times$~latitude)}} & \colhead{Number of vertical levels (top layer pressure)} & \colhead{Slab ocean depth (number of layers)}}
\startdata
ExoCAM  & 72~$\times$~46 & 51 (1~Pa) & 100~m (1)\\
\midrule
\addlinespace[0.5cm]
LMD-G   & 72~$\times$~46 & 40 (4~Pa) & 20~m (18) \\
\midrule
\addlinespace[0.5cm]
ROCKE-3D & 72~$\times$~46 & 40 (10~Pa) & 90~m (13) \\
\midrule
\addlinespace[0.5cm]
UM & 144~$\times$~90 & \makecell{Hab~1: 41 (4~Pa)\\Hab~2: 38 (13~Pa)} & 1~m (1) \\
\enddata
\end{deluxetable*}
The horizontal and vertical resolution of each model is given in Table~\ref{tab:grids}.
We refer to the Part I \citep{Turbet21_THAI} of this trilogy for a further description of dynamical cores and model grids.
Here we provide a brief description of the model components related to the moist physics.
Note that no model parameterizations were specifically adapted for the THAI intercomparison and thus the model configurations are close to what has been used in their recent single-model studies.
A comparison of moist physics parameterizations in THAI GCMs is given in Table~\ref{tab:models}.

\begin{longrotatetable}
\begin{deluxetable*}{cc|c|c|c|c}
\tablecaption{Summary of moist physics parameterizations in THAI GCMs
\label{tab:models}}
\tabletypesize{\scriptsize}
\tablehead{
\colhead{} & \colhead{Scheme components} & \colhead{ExoCAM} & \colhead{LMD-G} & \colhead{ROCKE-3D} & \colhead{UM}
}
\startdata
\addlinespace[0.25cm]
\multirow{8}{*}{\rotatebox[origin=c]{90}{Radiative transfer}} & Type of scheme & Two-stream, correlated-$k$ & Two-stream, correlated-$k$ & \multicolumn{2}{c}{Two-stream, correlated-$k$} \\
                                    & \makecell{SW spectral range\\\& number of bands} & \multirow{2}{*}{\SIrange{0.2}{1000}{\micro\m}; 28 bands} & \makecell{Hab~1: \SIrange{0.3}{5.1}{\micro\m}; 36 bands\\Hab~2: \SIrange{0.3}{5.1}{\micro\m}; 36 bands} & \multicolumn{2}{c}{\makecell{Hab~1: \SIrange{0.2}{20}{\micro\m}; 21 bands\\Hab~2: \SIrange{0.2}{20}{\micro\m}; 42 bands}} \\
                                    & \makecell{LW spectral range\\\& number of bands} & & \makecell{Hab~1: \SIrange{2.3}{1000}{\micro\m}; 38 bands\\Hab~2: \SIrange{2.3}{1000}{\micro\m}; 32 bands} & \multicolumn{2}{c}{\makecell{Hab~1:  \SIrange{3.3}{10,000}{\micro\m}; 12 bands\\Hab~2: \SIrange{3.3}{10,000}{\micro\m}; 17 bands}} \\
                                    & \makecell{Spectroscopy\\database} & HITRAN2004 & HITRAN2012 & \multicolumn{2}{c}{HITRAN2012} \\
                                    & Continuum & MT\_CKD v2.5 & MT\_CKD v2.5 & \multicolumn{2}{c}{MT\_CKD v3.0} \\
                                    & \makecell{Liquid particles\\effective radius} & \SI{14}{\micro\m} & \multirow{2}{*}{\makecell{Set by the CCN\\number and cloud water\\mixing ratio}} & \multicolumn{2}{c}{\makecell{Set by the CCN number\\and liquid water content;\\parameterization of\\thick averaging\\with Pade fits}} \\
                                    & \makecell{Ice particles\\effective radius} & Temperature-dependent & & \multicolumn{2}{c}{See \citet{Edwards07}} \\
                                    & \makecell{Subgrid-scale\\cloud variability} & MCICA & None & \multicolumn{2}{c}{Scaling factors} \\
\addlinespace[0.10cm]
\midrule
\addlinespace[0.10cm]
\multirow{6}{*}{\rotatebox[origin=c]{90}{Convection}} & Type of scheme & Mass-flux & Adjustment & Mass-flux & Mass-flux \\
                            & Subtypes & Deep, shallow & None & None & Deep, mid-level, shallow \\
                            & Trigger & \makecell{Conditional instability\\diagnosed by entraining\\ plume ascent} & Conditional instability & \makecell{Conditional instability\\ diagnosed by one\\undilute and one\\entraining plume} & \makecell{Conditional instability\\diagnosed by undilute\\parcel ascent} \\
                            & Closure & CAPE-based & None & \makecell{Mass flux reaching\\neutral buoyancy} & \makecell{CAPE-based\\and dependent\\on the vertical\\velocity for deep\\and mid-level\\convection, cloud-based\\for shallow\\convection} \\
                            & Downdrafts & \checkmark & $\times$ & \checkmark & \checkmark \\
                            & \makecell{Convective momentum\\transport} & \checkmark & \checkmark & \checkmark & \checkmark \\
\addlinespace[0.10cm]
\midrule
\addlinespace[0.10cm]
\multirow{7}{*}{\rotatebox[origin=c]{90}{\makecell{Large-scale clouds\\ \& precipitation}}} & \makecell{Treatment of\\cloud condensate} & Prognostic & Prognostic & Prognostic & Prognostic \\
                                    & \makecell{Treatment of\\cloud fraction} & Diagnostic & Diagnostic & Diagnostic & Prognostic \\
                                    & \makecell{Liquid/ice cloud\\particles separation} & Temperature-dependent & Temperature-dependent & Temperature-dependent & Process-based \\
                                    & \makecell{Supercooled\\water limit} & \SI{233.15}{\K} & \SI{255.15}{\K} & \SI{233.15}{\K} & \SI{233.15}{\K} \\
                                    & Column cloud fraction & \makecell{Maximum-random\\overlap} & Maximum overlap & \makecell{Maximum-random\\overlap} & \makecell{Exponential random\\overlap} \\
                                    & CCN number (\si{\per\m\cubed}) & Fixed (see text for details) & \makecell{\num{e7} for liquid;\\\num{e4} for ice particles} & Fixed (see text for details) & \makecell{\num{e8} for liquid;\\ variable for ice particles} \\
                                    & Precipitation & Diagnostic & Diagnostic & Diagnostic & Prognostic \\
\addlinespace[0.10cm]
\enddata
\end{deluxetable*}
\end{longrotatetable}

\subsubsection{Radiative transfer} \label{sec:rt}
While the main description of radiative transfer (RT) parameterizations is given in THAI Part I \citep{Turbet21_THAI}, we provide additional details in this section on how these parameterizations deal with water vapor and cloud particles.
The RT schemes used by THAI GCMs share a number of characteristics: they are based on the two-stream correlated-$k$ approach; the HITRAN database is used for spectroscopic data, MT\_CKD formalism is used for continuum absorption\footnote{ExoCAM and LMD-G use the MT\_CKD v2.5, while ROCKE-3D and the UM use MT\_CKD v3.0. The newer version leads a better treatment of \ce{H2O} shortwave absorption in the near-infrared, which can be particularly important to \ce{H2O}-rich climates \citep{yang_differences_2016,Wolf2022}}.
They take into account the scattering effects of the atmosphere (Rayleigh scattering) and \ce{H2O} clouds (Mie scattering).
There are however, several key differences among the schemes, as outlined below.

ExoCAM simulations are performed with ExoRT, whose oldest and most published version is used here; it is referred to as \texttt{n28archean} \citep{Wolf&Toon2013,yang_differences_2016}.
This version has 28 bands covering the spectral range between 0.2 and \SI{1000}{\micro\m}.
Note that subsequent intercomparison studies have found this version of the radiative transfer to overestimate the greenhouse effect from pure \ce{CO2} atmospheres \citep{Wolf2022}, and overestimate the near-infrared absorption of \ce{H2O} for warm moist atmospheres irradiated by M-dwarf stars \citep{yang_differences_2016}.
Thus, particularly for Hab~2, we would expect this version of radiative transfer to have a bias towards producing a climate that is too warm based on clear-sky radiative transfer considerations only.
Later iterations of ExoRT have corrected these issues, primarily by increasing the spectral resolution in the near-infrared and by using a newer version of MT\_CKD continuum, as is discussed in \citet{Kopparapu2017} and \citet{Wolf2022}.
Liquid and ice cloud particles are treated as Mie scattering particles with refractive indices taken from \citet{Segelstein1981} and \citet{WarrenBrandt2008}.
Liquid cloud particle effective radii are fixed at \SI{14}{\micro\m}, while ice crystal sizes are allowed to vary dependent on the air temperature \citep[see][for more detail on the particle sizes and their implications for observations]{Fauchez21_THAI}.
Note that the cloud condensation nuclei (CCN) number used for the effective radii of cloud particles in ExoRT does not affect particle sizes in the large-scale cloud parameterization (see below).
The radiative effect of cloud overlap is treated using the Monte Carlo Independent Column Approximation while assuming maximum-random overlap \citep{Pincus2003, Barker2008}.

In LMD-G's RT module \citep{Wordsworth:2011}, the correlated-$k$ coefficients were built using between 32 and 38 spectral bands in the thermal infrared (from 2.3 to \SI{1000}{\micro\m}) and between 32 and 36 spectral bands in the visible domain (from 0.3 to \SI{5.1}{\micro\m}), depending on the atmospheric composition.
Mie scattering by cloud particles is parameterized following \citet{Hansen:1974}, and their refractive indices cloud particles are taken from \citet{Warren1984}.
The mean cloud particle radius is calculated from i) the CCN number and ii) the amount of condensed \ce{H2O} \citep[eq.~3 in][]{Turbet:2020}.
As described in \citet{Charnay:2013}, radiative transfer is computed twice: in a clear sky column, and in a cloudy sky column whose area is equal to the total cloud cover.

Both ROCKE-3D and the UM share a common RT module SOCRATES\footnote{\url{https://code.metoffice.gov.uk/trac/socrates}}, which is an open-source model developed by the Met Office \citep{Edwards96,Manners21}.
SOCRATES is highly flexible.
Spectral files, created for both shortwave (\SIrange{0.2}{20}{\micro\m}) and longwave (\SIrange{3.33}{10,000}{\micro\m}) streams separately, contain all necessary information to define the radiative transfer problem and are tailored to specific atmospheric compositions and stellar spectra, then input to the GCMs at run-time.
For both Hab~1 and Hab~2 simulations, the UM and ROCKE-3D use identical pairs of spectral files\footnote{Available at \url{https://portal.nccs.nasa.gov/GISS_modelE/ROCKE-3D/spectral_files/}}.
For Hab~1, both models use a spectral file tailored to present-day Earth-like conditions having 21 shortwave bins (\texttt{sp\_sw\_21\_dsa}) and 12 longwave bins (\texttt{sp\_lw\_12\_dsa}).
For Hab~2, both models use a spectral file with 42 shortwave (\texttt{sp\_sw\_42\_dsa\_mars}) and 17 longwave (\texttt{sp\_lw\_17\_dsa\_mars}) bands tailored to a \ce{CO2}-dominated atmosphere.
The parameterization of cloud droplets uses the method of ``thick averaging'' \citep{Edwards96} and Pad\'{e} fits for the variation with effective radius.
Ice crystals treatment uses the parameterization of \citet{Edwards07}.
The CCN number is treated consistently between the RT and cloud modules.
The subgrid-scale cloud inhomogeneity is taken into account using scaling factors, which depend on whether the cloud is stratiform or convective and consists of liquid or frozen condensate.

\subsubsection{Moist convection} \label{sec:conv}
To account for the subgrid-scale transport of heat and moisture by cumulus clouds, GCMs employ convection parameterizations.

In ExoCAM, deep penetrating moist convection is parameterized following \citet{Zhang95} with modifications to the numerical solver to improve numerical stability for warm climates \citep{Wolf2015}.
This mass-flux scheme represents deep convection using a plume ensemble approach, triggering updrafts and downdrafts by conditional instability, which is diagnosed by entraining plume ascent.
The parameterization is closed, i.e. the mass fluxes are determined as a function of the rate at which convective available potential energy (CAPE) is consumed.
A separate parameterization for shallow convection (i.e. between 3 or fewer layers) is treated following \citet{hack:1994}.

In contrast to the other three GCMs, LMD-G employs a convective adjustment scheme, which is a conceptually simpler parameterization, but does not represent the full complexity of subgrid-scale processes associated with moist convection.
This parameterization is based on the concept of radiative-convective equilibrium and lets convection rapidly mix heat and moisture in the vertical, effectively relaxing an unstable temperature profile to a reference state.
In other words, this convective adjustment scheme tries to maintain a specified moist adiabatic lapse rate in the atmospheric column.
Note that the Manabe-Wetherald scheme \citep{Manabe_Wetherald1967} is used instead of the more common Betts-Miller, because it is more robust for a wide range of pressures, though at the cost of giving enhanced precipitation near the substellar point \citep{Charnay:2013}.

Like ExoCAM, ROCKE-3D accounts for cumulus convection via the mass-flux approach \citep{Schmidt2014}.
The scheme is triggered by conditional instability calculated using two plumes: undilute and dilute (entraining).
This allows for the representation of different types of cumulus clouds without invoking a separate shallow convection parameterization.
The scheme's closure is based on the assumption of neutral buoyancy achieved by the mass flux at the cloud base. 
Relative to its parent code, the Earth GCM ModelE 2, ROCKE-3D includes a few modifications to the convection scheme: e.g. a stronger entrainment rate, more vigorous vertical transport of condensed water, and a relaxed limit for the convection top pressure \citep{Way2017}.

The UM also uses a mass-flux convection scheme, which is based on \citet{Gregory90} and developed further to improve the representation of downdrafts and convective momentum transport \citep{Walters19}.
The convection is triggered by the surface buoyancy flux diagnosed by the boundary-layer scheme and an undilute parcel ascent to the level of neutral buoyancy.
The type of convection is then classified as shallow, mid-level or deep, depending on the height of the level of neutral buoyancy relative to the freezing level and the level of strong vertical updrafts.
For deep and mid-level convection, the UM uses a CAPE-based closure with a dependency on the grid-scale or vertical velocity; for shallow convection a cloud-based closure is used.

\subsubsection{Clouds and precipitation} \label{sec:clouds}
While cumulus physics in GCMs are handled by the convection scheme, grid-scale cloud processes are the responsibility of the large-scale cloud and precipitation schemes.
Depending on the complexity of the model, cloud-related processes directly impact the state of the atmosphere by changing its radiative properties and by releasing latent heat of condensation or freezing.
There is a spectrum of cloud parameterizations, from purely diagnostic \citep[e.g.][]{Smith90,Liu21}, to mixed prognostic-diagnostic \citep[e.g.][]{Sundqvist78}, to fully prognostic \citep[e.g.][]{Wilson08}, as well as statistical schemes based on various moments of subgrid-scale variability.
While diagnostic schemes offer transparency and simplicity of representing clouds in a GCM, the advantage of the prognostic approach is more physically based evolution of clouds, including the web of complex interactions between vapor and condensate phases, and better treatment of cloud optical properties by the RT code.
As an example, a prognostic cloud scheme makes it possible for cloud to be transported away from where it formed, as in the case of anvils detrained from convection that can persist after the convection itself has ceased.
Three THAI GCMs rely on Sundqvist-type prognostic-condensate/diagnostic-cloud schemes, while the UM uses a fully prognostic approach.

ExoCAM relies on the CAM4 moist physics package \citep{Neale2010} to control water vapor, liquid cloud, and ice cloud condensate fields via prognostic bulk microphysical parameterizations.
Cloud fraction is diagnosed in three categories: marine stratus clouds dependent on the temperature profiles; convective clouds defined by the convective mass-flux; and layered clouds forming at a sufficient relative humidity threshold.
The total cloud fraction within each grid box is diagnosed via the maximum overlap assumption. 
ExoCAM cloud scheme uses a fixed cloud droplet number concentration, equal to 1.5~$\times10^{8}$~$m^{-3}$, which is used to compute the conversion of liquid water to rain \citep{Neale2010}.
Precipitation is treated diagnostically and can be in a form of rain or snow, or a mixture of both.

In LMD-G, the large-scale condensation is based on the prognostic equations of \citet{Letreut1991} that track the evolution of the total cloud condensate, which is split into the liquid or ice phase according to eq.~2 of \citet{Charnay:2013}.
Liquid cloud droplets are allowed to exist in the supercooled state down to the limit of \SI{255.15}{\K}, below which clouds are only composed of ice crystals.
Cloud fraction is diagnosed from relative humidity.
The total cloud cover of each vertical atmospheric column is taken equal to the partial cover of the optically thickest cloud of the column.
LMD-G then assumes that each individual cloud cover is equal to this total cloud cover with a maximum overlap \citep{Charnay:2013} --- equivalent to the ``LMDG\_max'' simulation in \citet{Yang2019}.
The CCN number is set to \num{e7} and \SI{e4}{\per\kg} for liquid and ice cloud particles, respectively, following \citet{Turbet:2020}.
Water precipitation is divided into rainfall \citep{Boucher1995} and snowfall, and is computed as in \citet{Charnay:2013}.
For both cases, the LMD-G assumes that precipitation is instantaneous (i.e. it goes directly to the surface), though it can evaporate while falling through subsaturated layers, thus producing water vapor and latent heat cooling.
The precipitation re-evaporation scheme is based on \citet{Gregory95_consistent} and is discussed in more detail in \citet{Charnay:2013}.

ROCKE-3D uses a prognostic cloud water approach with diagnostic cloud fraction \citep{Schmidt2014,Way2017}.
Like in ExoCAM and LMD-G, a single cloud water variable is predicted; its phase is then diagnosed from temperature and corrected for precipitation from upper model layers.
Note, however, that the cloud condensate is not advected by resolved winds or turbulence.
Cloud volume fraction is a function of relative humidity, with a different threshold for the boundary layer and free troposphere.
The column cloud fraction is computed using maximum-random overlap.
The cloud particle number is fixed over the ocean surface (and thus for the whole atmosphere in Hab~1 and Hab~2 simulations) and is equal to \SI{\approx 6e7}{\per\m\cubed} for liquid clouds and to \SI{\approx 6e4}{\per\m\cubed} for ice clouds \citep{DelGenio96_prognostic}.
Precipitation is treated diagnostically.

The UM represents large-scale clouds via the Prognostic Cloud fraction and Prognostic Condensate (PC2) scheme \citep{Wilson08} along with additional parameterizations of cloud erosion and critical relative humidity.
The scheme includes three prognostic mixing ratios (water vapor, liquid and ice) and three prognostic cloud fractions (liquid, ice and mixed-phase).
The prognostic variables are updated by process-driven increments, and all cloud types are handled by single scheme, with the assumption of minimum overlap of phases.
The column cloud fraction is a product of exponential random vertical overlap.
The CCN number is fixed at \SI{e8}{\per\m\cubed} for liquid droplets and is variable for ice crystals.
Precipitation is formed according to a single-moment scheme based on \citet{Wilson99}.
The warm rain is based on \citet{Boutle14b} and includes prognostic rain, allowing for 3D transport and sub-grid variability of precipitation.
Rain-rate-dependent particle size and fall velocities allow the UM to represent better sedimentation and evaporation of small droplets \citep{Walters19}.

\section{Results} \label{sec:results}
For each case, Hab~1 (Sec.~\ref{sec:hab1}) and Hab~2 (Sec.~\ref{sec:hab2}), we present results from the four GCMs, comparing and contrasting them against each other and previous studies.
We report the following key aspects of the climate in our numerical simulations.
We first discuss radiation fluxes at the top of the atmosphere and at the planet's surface, following up with the analysis of temperature and humidity profiles in each of the GCMs.
We then discuss the cloud amount and its spatial distribution.
We then move on to comparing the GCMs with respect to the mean large-scale atmospheric circulation.
We conclude the analysis of each of the cases by analysing the time variability of cloudiness at the planet's limb.

\begin{deluxetable*}{lcccccccccc}
\renewcommand{\arraystretch}{1.25}
\tablecaption{Global mean surface temperature (T\textsubscript{s}, \si{\K}) global mean top-of-atmosphere shortwave (CRE\textsubscript{SW}), longwave (CRE\textsubscript{LW}), and net cloud radiative effect (CRE, \si{\watt\per\square\metre}), day side net CRE (CRE\textsubscript{day}), night side net CRE (CRE\textsubscript{night}), planetary albedo ($\alpha_\text{p}$), all-sky (G\textsubscript{as}) and clear-sky (G\textsubscript{cs}) greenhouse effect (\si{\K}), water vapor volume mixing ratio at the \SI{1}{\hecto\pascal} level (VMR\textsubscript{\SI{1}{\hecto\pascal}}, \si{\mol\per\mol}) in Hab~1 and Hab~2 simulations.\label{tab:glob_diag}}
\tablewidth{0pt}
\tablehead{
GCM & \multicolumn{10}{c}{Hab~1} \\
{} & T\textsubscript{s} & CRE\textsubscript{SW} & CRE\textsubscript{LW} &   CRE & CRE\textsubscript{day} & CRE\textsubscript{night} & $\alpha_\text{p}$ & G\textsubscript{as} & G\textsubscript{cs} & VMR\textsubscript{\SI{1}{\hecto\pascal}}
}
\startdata
ExoCAM  & 245.6 &    -37.9 &     15.5 & -22.4 & -46.0 & -0.3 & 0.24 &   10.3 &    5.8 &  \num{4.6e-07} \\
LMD-G    & 242.2 &    -21.8 &      5.1 & -16.8 & -33.2 & -1.2 & 0.22 &    6.0 &    4.8 &  \num{5.8e-08} \\
ROCKE-3D & 244.0 &    -44.6 &      9.3 & -35.3 & -70.9 &  0.3 & 0.27 &    9.3 &    6.7 &  \num{5.1e-07} \\
UM      & 231.6 &    -48.1 &     10.1 & -38.0 & -70.9 & -5.0 & 0.28 &    2.6 &    0.0 &  \num{5.1e-07} \\
\midrule
{} & \multicolumn{10}{c}{Hab~2} \\
ExoCAM  & 295.2 &    -22.0 &     13.2 &  -8.8 & -26.9 &  8.0 & 0.15 &   53.0 &   48.9 &  \num{1.5e-05} \\
LMD-G    & 270.9 &    -32.3 &      3.2 & -29.1 & -60.9 &  1.0 & 0.21 &   34.3 &   33.2 &  \num{3.8e-06} \\
ROCKE-3D & 284.0 &    -35.0 &      8.6 & -26.4 & -58.7 &  5.9 & 0.19 &   44.7 &   41.9 &  \num{4.5e-05} \\
UM      & 280.4 &    -27.3 &      8.3 & -19.0 & -41.7 &  3.6 & 0.16 &   39.1 &   36.5 &  \num{9.2e-06} \\
\enddata
\end{deluxetable*}

\subsection{Global climate in Hab~1 simulations} \label{sec:hab1}
\subsubsection{Radiation fluxes and thermodynamic profiles} \label{sec:hab1_thermo}
At a planetary scale, the climate of TRAPPIST-1e is determined by the amount of the shortwave stellar radiation absorbed at the top of the atmosphere (TOA) on the day side and the amount of longwave radiation lost to space.
In this study, as in many other GCM studies, the shortwave (SW) and longwave (LW) streams of radiation refer to the radiation emitted by the host star and by the planet, respectively.
In all four THAI GCMs, there is an overlap in wavelengths of the SW and LW streams, as noted in Table~\ref{tab:models}.
The clear-sky absorbed radiation peaks at the substellar point, as in the dry experiments \citep[see Part I,][]{Turbet21_THAI}, but the cloud cover reflects a large part of shortwave radiation back to space.
The impact of clouds on radiation is evident in all four GCMs and can be quantified as the cloud radiative effect $\textrm{CRE}=F_{\textrm{clear-sky}} - F_{\textrm{all-sky}}$, where $F_{\textrm{clear-sky}}$ is the upward TOA radiative flux assuming clouds are absent\footnote{The clear-sky radiative fluxes are calculated within the radiative transfer schemes in each GCM and provided as standard output in the THAI simulations.} and $F_{\textrm{all-sky}}$ is the actual upward TOA radiative flux \citep[e.g.][]{Ceppi17}.
Here the CRE is averaged over the whole planet.
The shortwave CRE, CRE\textsubscript{SW}, is negative in the THAI simulations (Table~\ref{tab:glob_diag}), because clouds have a higher albedo than the planet's surface and thus reduce the amount of absorbed energy.
In Hab~1, the strongest CRE\textsubscript{SW} is exhibited by the UM (\SI{-48.1}{\watt\per\m\squared}) and is close to the average value for the modern Earth's climate, \SI{-45}{\watt\per\m\squared} \citep[e.g.][]{Henderson13}.
Even though the UM does not have the highest cloud amount among the GCMs on average, the cloud cover is concentrated and is the thickest on the day side of the planet (see Sec.~\ref{sec:hab1_clouds}).
Hab~1 climate CRE\textsubscript{SW} estimates (multiplied by 2 because there is no CRE\textsubscript{SW} on the night side) are also close to the typical values observed over the warm pools of the tropical Pacific and Indian Oceans on Earth, where CRE\textsubscript{SW} is $60-80~W~m^{-2}$ \citep{Wall19_net}.
The weakest CRE\textsubscript{SW} is exhibited by LMD-G (\SI{-21.8}{\watt\per\m\squared}) and is smaller by more than a half of that found for simulations using the other GCMs.
This is primarily due to a reduced cloud fraction and small cloud water content produced by LMD-G on the day side (Fig.~\ref{fig:hab1_cloud_cross} and \ref{fig:hab1_cloud_maps}).
The second reason is the size of cloud particles, which is set to a constant or parameterized depending on variables such as temperature and cloud condensation nuclei number \citep[][see also Sec.~\ref{sec:clouds} for details]{Turbet:2020}.
The effective radius of cloud droplets and crystals tends to be the largest in LMD-G compared to that in the other three models (not shown).
Larger cloud particles result in a decrease of the shortwave optical depth of clouds, which leads to a decrease in the reflected radiation flux at TOA, thus making the CRE\textsubscript{SW} less negative.

\begin{figure*}
\includegraphics[width=\textwidth]{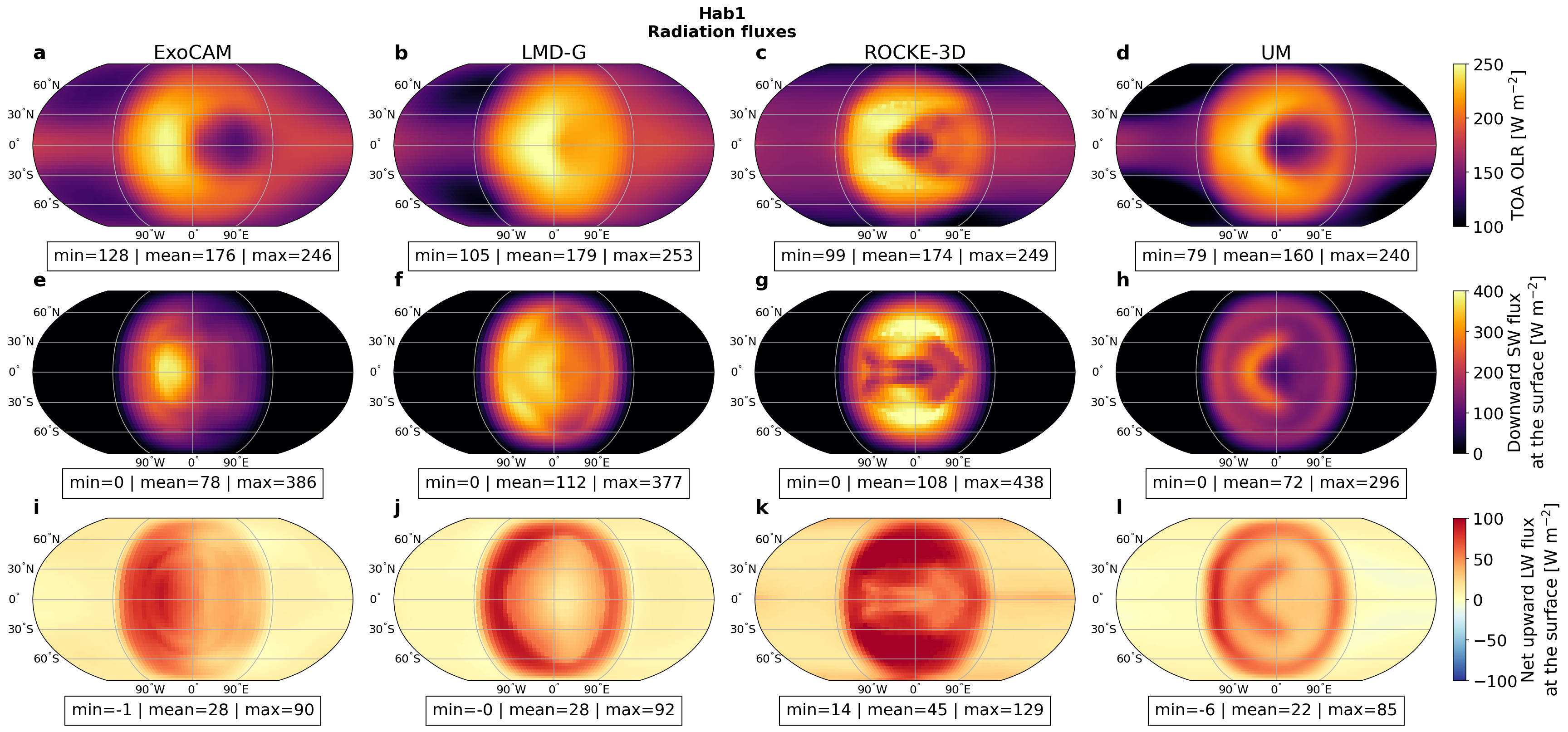}
\caption{Key radiation fluxes in Hab~1 simulations. (a--d) top-of-the-atmosphere outgoing longwave radiation (TOA OLR, \si{\watt\per\m\squared}), (e--h) downward shortwave radiation flux at the surface (\si{\watt\per\m\squared}), (i--l) net upward longwave radiation flux at the surface (\si{\watt\per\m\squared}). \label{fig:hab1_rad_flux}}
\end{figure*}

The cloud cover, produced by the interaction between the convective (Sec.~\ref{sec:conv}) and large-scale cloud (Sec.~\ref{sec:clouds}) parameterizations and shaped by the atmospheric circulation, leaves a distinct imprint in the TOA radiation fluxes and shifts the shortwave absorption maximum away from the substellar point in all GCMs (not shown, but the pattern is similar to that shown in Fig.~\ref{fig:hab1_rad_flux}e--h).
The resulting planetary albedo repeats the trend in CRE\textsubscript{SW} (Table~\ref{tab:glob_diag}), with LMD-G having the albedo of about 0.22, followed by ExoCAM (0.24), then ROCKE-3D (0.27) and the UM (0.28).

While the strongest TOA longwave emission is on the day side (exceeding \SI{240}{\watt\per\m\squared}), a significant portion of it is on the night-side too, especially in low latitudes of the eastern hemisphere (Fig.~\ref{fig:hab1_rad_flux}a--d).
Due to high-altitude clouds, the longwave emission has a pronounced minimum at and to the east of the substellar point, especially in ExoCAM, ROCKE-3D and the UM.
The global longwave CRE, CRE\textsubscript{LW}, is positive, because high cloud cover reduces the amount of outgoing thermal radiation; reaches \SI{15.5}{\watt\per\m\squared} in the ExoCAM simulation, dropping by a third for ROCKE-3D and the UM and then even more for LMD-G.
Compared to \SI{27}{\watt\per\m\squared} for Earth \citep{Henderson13}, CRE\textsubscript{LW} in the Hab~1 simulations is smaller, which is explained chiefly by the low CRE\textsubscript{LW} on the night side, where the low altitude of clouds coincides with the highest temperature due to the boundary layer thermal inversion \citep{Sergeev2020}.
The effective radii of cloud particles offers a small contribution to the inter-model spread of CRE\textsubscript{LW}, due to its smaller effect on the longwave fluxes \citep[e.g.][]{Abbot14}.

The overall effect of clouds on TRAPPIST-1e is cooling, confirming the earlier theory of clouds stabilizing the climate of tidally locked terrestrial exoplanets and expanding the inner edge of the habitable zone \citep{Joshi03,Yang2013,Yang:2014,Kopparapu2016}.
The magnitude of the net CRE is the largest in the UM (\SI{-38.0}{\watt\per\m\squared}) and ROCKE-3D (\SI{-35.3}{\watt\per\m\squared}), diminishing by almost a half in ExoCAM (\SI{-22.4}{\watt\per\m\squared}) and lower still in LMD-G (\SI{-16.8}{\watt\per\m\squared}; Table~\ref{tab:glob_diag}).
These numbers summarize that the radiative influence of clouds on the Hab~1 climate is the strongest in the UM and the weakest in the LMD-G.
The tendency of the LMD-G to produce weaker CRE than that in other GCMs, mostly due to its shortwave part, was also reported by \citet{Yang2019}, although the amplitude of CRE is significantly lower in our experiments due to lower stellar flux and due to faster rotation of the planet \citep{Kopparapu2016}.
To sum up, the models have different CRE because of the distribution of clouds, especially on the day side of the planet, and because of the size of cloud particles, which affects primarily the shortwave radiation fluxes.

\begin{figure*}
\includegraphics[width=\textwidth]{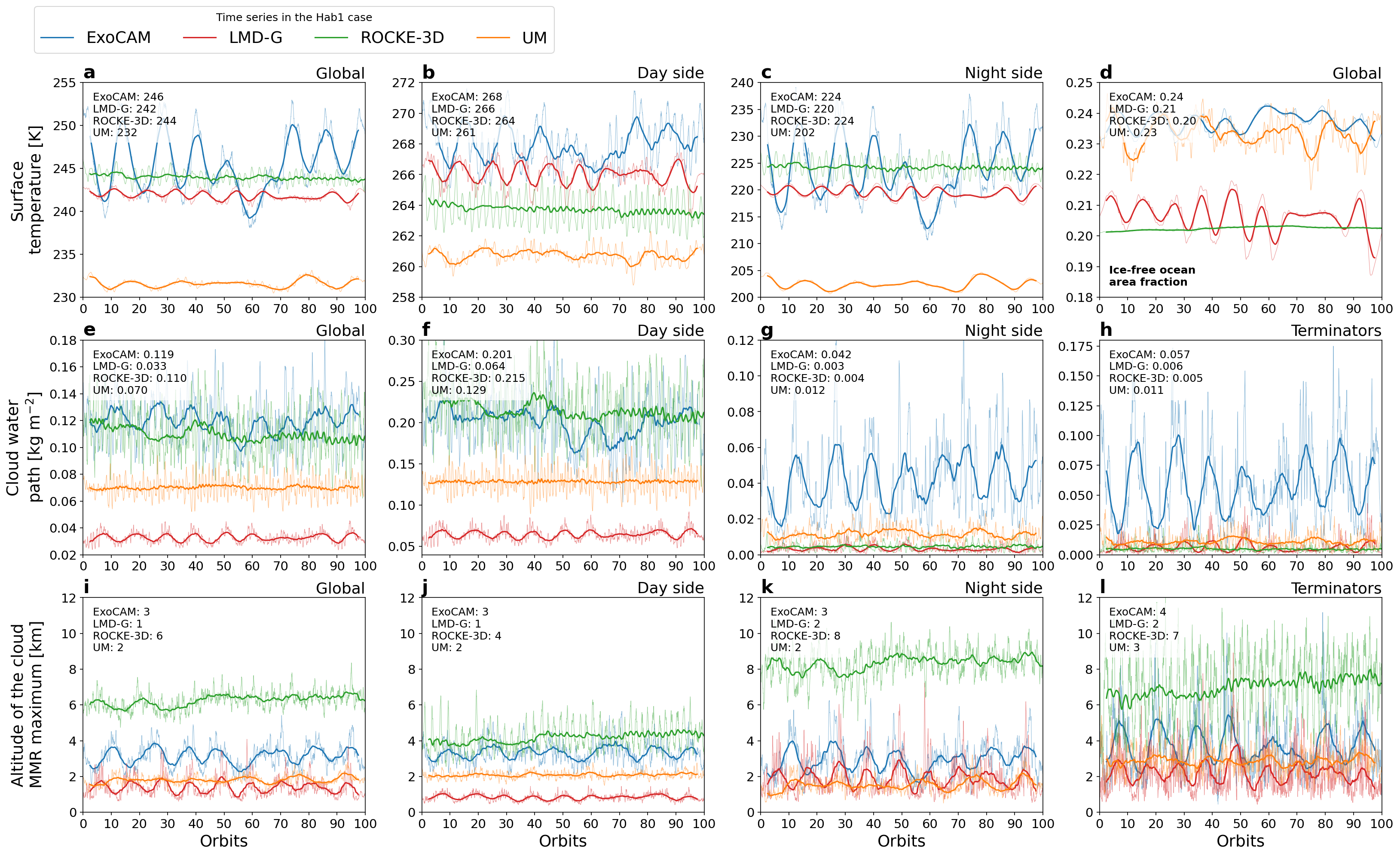}
\caption{Hab~1 results: time variability of (a--c) surface temperature (\si{\kelvin}), (d) ice-free ocean area fraction, (e--h) cloud water path (\si{\kg\per\m\squared}) and (i--l) the altitude of the cloud mass mixing ratio (MMR) maximum (\si{\km}). Each column shows (a, d, e, i) global mean, (b, f, j) day-side mean, (c, g, k) night-side mean and (h, l) terminator mean. Thin lines show the raw data, thick lines show a 5-orbit rolling mean. \label{fig:hab1_tseries}}
\end{figure*}

Due to the reflection and scattering by the day-side clouds, the stellar radiation flux reaching the planet's surface is unevenly distributed around the substellar point (Fig.~\ref{fig:hab1_rad_flux}e--h).
Its maximum, located broadly to the west of the substellar longitude, reaches values as high as \SI{438}{\watt\per\m\squared} in the ROCKE-3D case, while it is only \SI{296}{\watt\per\m\squared} in the UM case.
The arrow-shaped distribution of clouds in the ROCKE-3D simulation (Fig.~\ref{fig:hab1_cloud_maps}c) shields the equatorial region from the stellar radiation, instead allowing the flux to reach its maximum at \ang{\approx 45} latitude. 
This cloud pattern, and the radiation flux pattern as a consequence, is due to the dynamical regime in ROCKE-3D being distinctly different from that in the other 3 GCMs (see Sec.~\ref{sec:hab1_circ}).
Namely, the ROCKE-3D simulation is in a rapid-rotator regime, in which the meridional circulation and convective activity sustaining the cloud cover is confined to a narrow equatorial band, similar to the simulations of a planet orbiting a late M-dwarf in \citet{Komacek&Abbot2019}.
In terms of the global average of the surface stellar flux, GCMs fall into two groups: those with the flux of about \SI{75}{\watt\per\m\squared} (ExoCAM and UM) and those with the flux of about \SI{50}{\percent} higher (LMD-G and ROCKE-3D).
This dichotomy is explained by the overall low amount of clouds in the LMD-G case and the concentrated distribution of clouds in the equatorial band in the ROCKE-3D case (see Sec.~\ref{sec:hab1_clouds}).
The ice-free ocean surface, occupying almost a half of the day-side area (Fig.~\ref{fig:hab1_tseries}d), absorbs \SI{94}{\percent} of the available stellar radiation, while the ``ice-covered'' periphery of the day side absorbs \SI{75}{\percent} (due to the fixed albedo, see Sec.~\ref{sec:setup}).  

\begin{figure*}
    \centering
    \includegraphics[width=\textwidth]{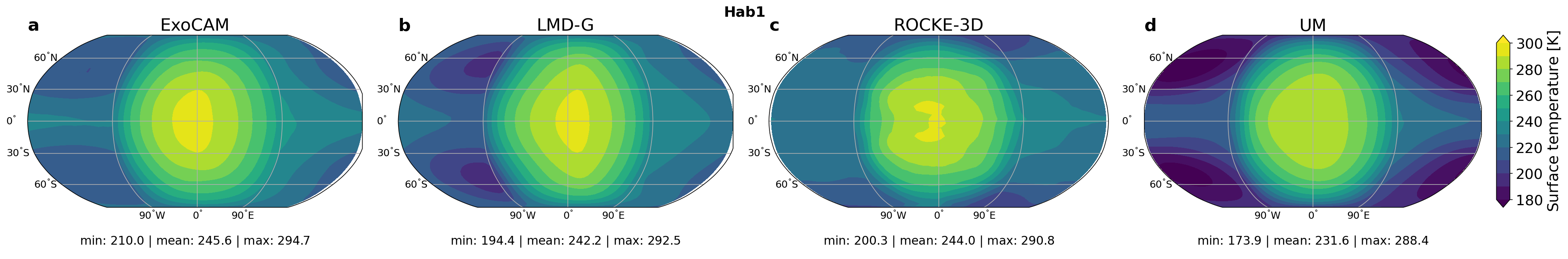}
    \caption{Hab~1 results: time mean surface temperature (\si{\K}) in (a) ExoCAM, (b) LMD-G, (c) ROCKE-3D, and (d) the UM.
    \label{fig:hab1_t_sfc_map}}
\end{figure*}

\begin{figure}
\includegraphics[width=0.45\textwidth]{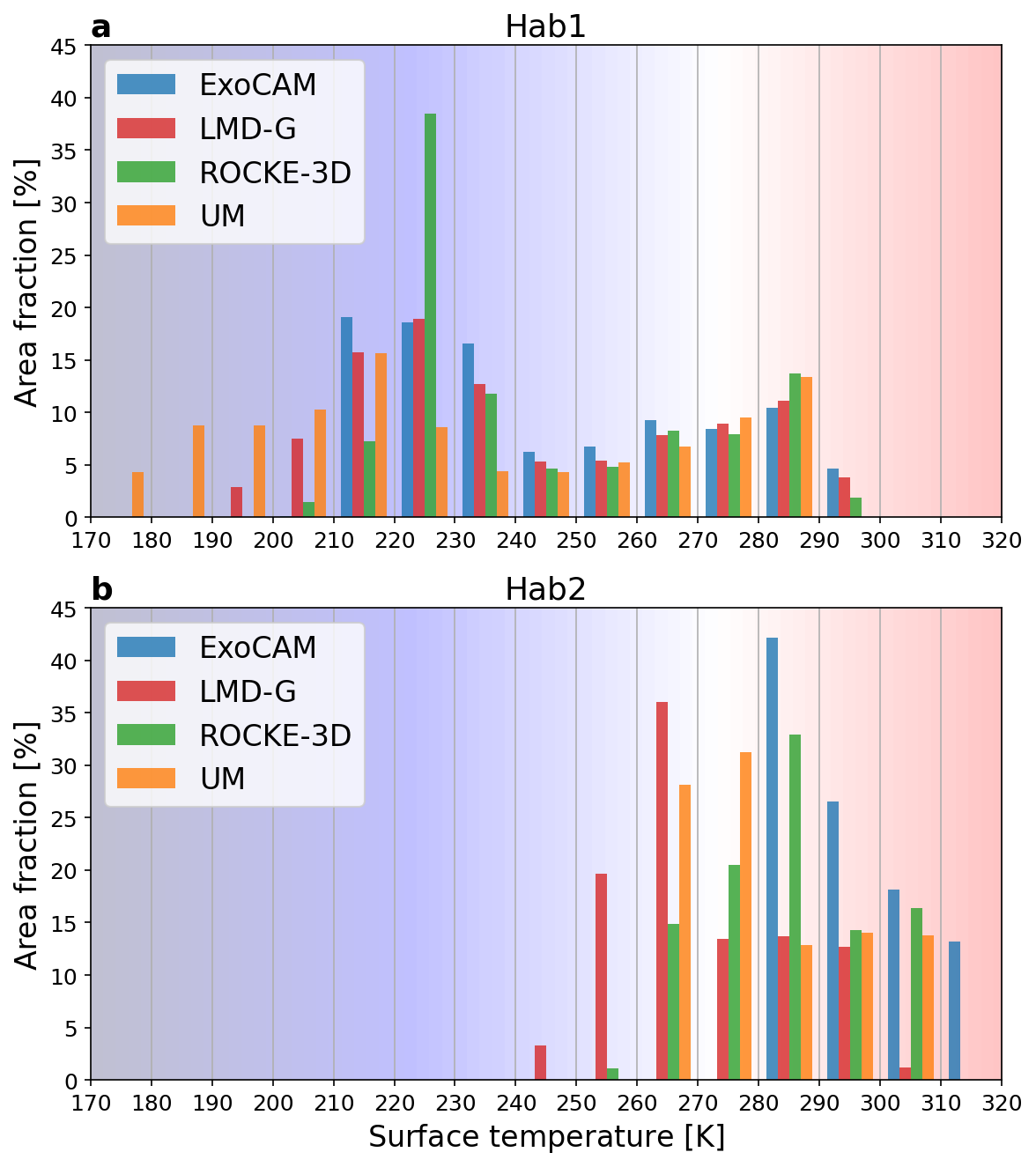}
\caption{Surface area fraction with corresponding time mean temperature binned every \SI{10}{\K} in (a) Hab~1 and (b) Hab~2. \label{fig:hist_t_sfc}}
\end{figure}

\begin{figure*}
\includegraphics[width=\textwidth]{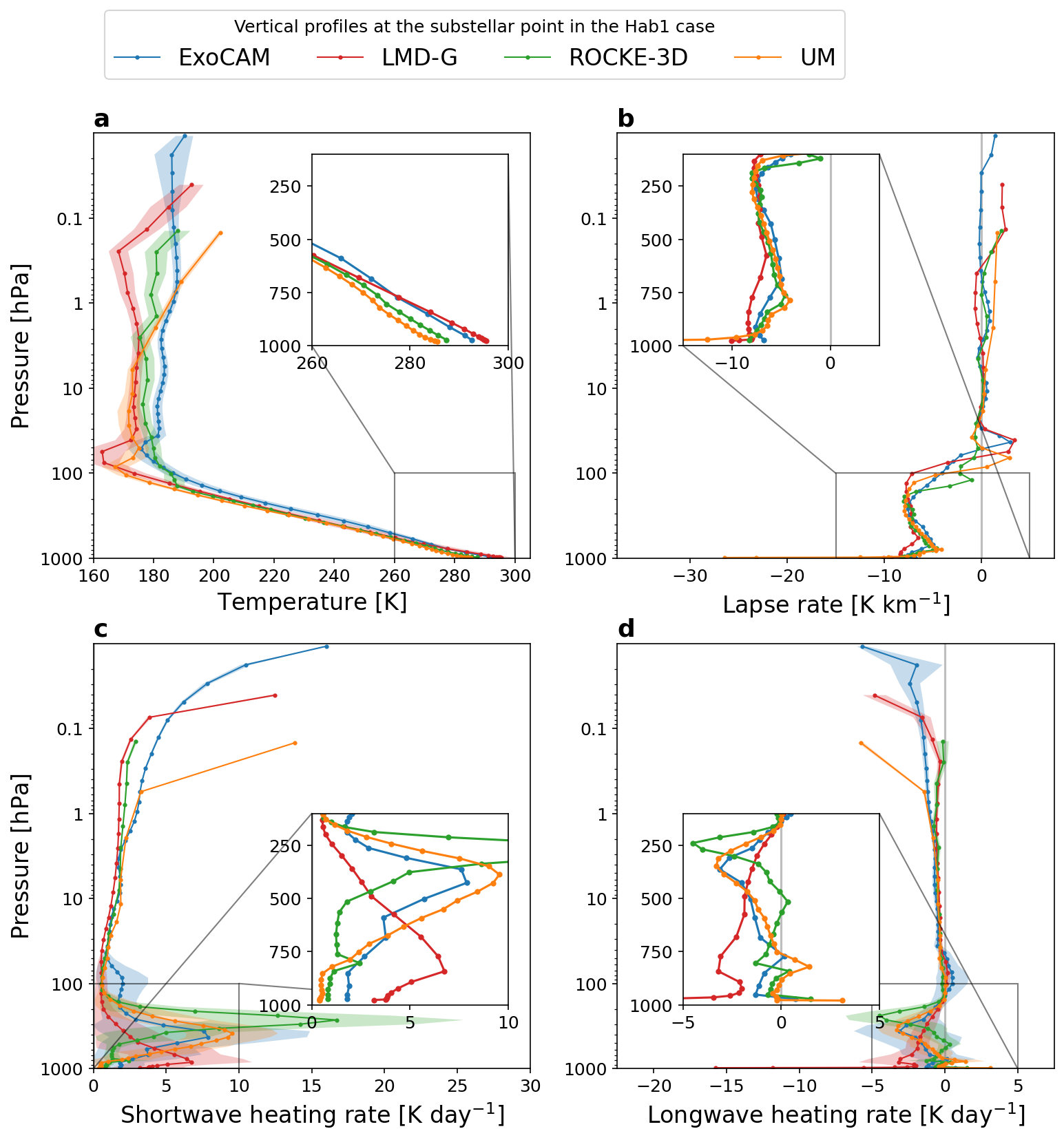}
\caption{Hab~1 results: substellar vertical profiles of (a) air temperature (\si{\kelvin}), (b) lapse rate (\si{\kelvin\per\km}), (c) SW heating rate (\si{\kelvin\per\day}), and (d) LW heating rate (\si{\kelvin\per\day}).
Time mean values are shown by solid lines, the 1-$\sigma$ deviation is shown by shading.
\label{fig:hab1_substellar}}
\end{figure*}

\begin{figure*}
\includegraphics[width=\textwidth]{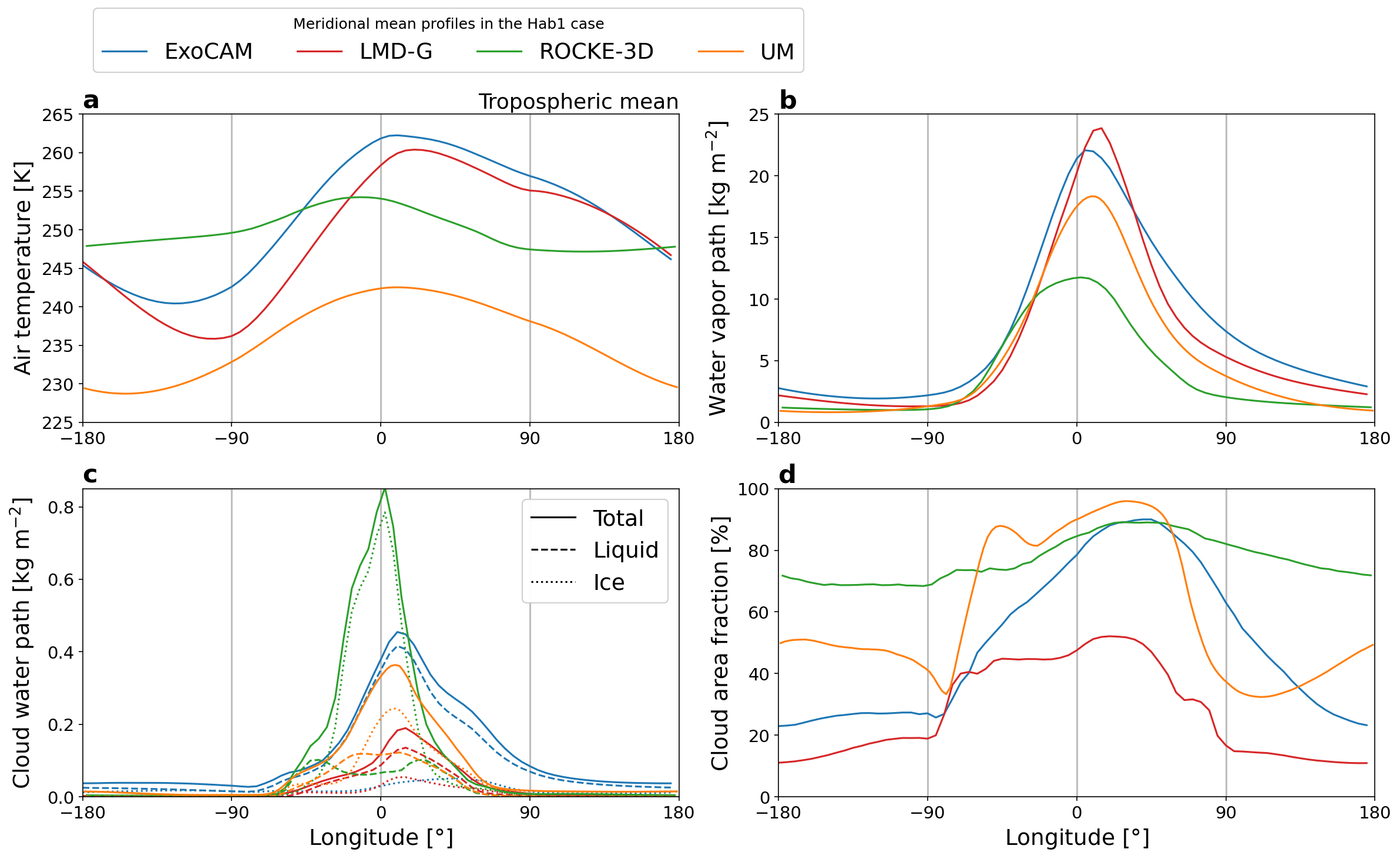}
\caption{Hab~1 results: meridional mean profiles of (a) the mass-weighted tropospheric mean of air temperature (\si{\kelvin}), (b) water vapor path (\si{\kg\per\m\squared}), (c) cloud water path (\si{\kg\per\m\squared}): total (solid lines), liquid (dashed lines), ice (dotted lines), and (d) cloud area fraction (\si{\percent}). Here, the troposphere's boundary is defined by the lapse rate decreasing below \SI{2}{\K\per\km} in the upper atmosphere. \label{fig:hab1_mm_prof}}
\end{figure*}

Our models predict mean global surface temperatures of \SIrange{\approx 232}{246}{\K}  (Table~\ref{tab:glob_diag}, Fig.~\ref{fig:hab1_tseries}a, Fig.~\ref{fig:hab1_t_sfc_map}), but the inter-model differences in temperature do not simply follow the differences between model planetary albedos.
Whilst ExoCAM, LMD-G and ROCKE-3D have similar estimates of the mean temperature as well as the temperature extrema, the UM produces a consistently colder climate (Fig.~\ref{fig:hab1_t_sfc_map}, \ref{fig:hist_t_sfc}a).
Because this difference is mostly due to the night side temperatures of the surface below the stationary cyclonic mid-latitude gyres (Fig.~\ref{fig:hab1_tseries}d, see also the mean circulation in Fig.~\ref{fig:hab1_rot_div}a--d), the day-night contrast in the UM is larger by \SI{\approx 20}{\K} than that in the other three models.
The surface temperature in ExoCAM is somewhat higher than that in LMD-G, which contradicts previous studies in which LMD-G was much warmer than CAM-family models \citep{Yang2019}.
The opposite outcome found here is not likely due to model heritage issues, because the model configurations changed minimally since that intercomparison was conducted.
Rather, the opposite outcome is likely a function of natural model behaviors arising from two different circulation regimes.
\citet{Yang2019}'s intercomparison studied slow rotating planets (60 day periods) receiving the same stellar insolation as modern Earth but from a \SI{3400}{\K} star with a blackbody radiation spectrum.
Here, TRAPPIST-1e receives less flux but from a much redder star, and perhaps most importantly, its dramatically faster rotation rate shifts the atmospheric circulation into a distinct dynamical state \citep[e.g.][]{Haqq_Misra2018,Sergeev2020} compared to the planets studied by \citet{Yang2019}.
Therefore it is perhaps unsurprising that the complex coupling between atmospheric circulation, cloud and radiative processes yields non-linear model sensitivities when making changes to fundamental planetary configuration.
This finding is another incentive to perform model intercomparisons for other planetary configurations, as will be done within the CUISINES framework \citep{Fauchez2021_THAI_workshop}.

The day side surface temperature is, on the other hand, relatively similar in all models (Fig.~\ref{fig:hab1_t_sfc_map}).
This is perhaps unsurprising given that the day side is under a constant forcing from the stellar radiation flux, whilst the night-side temperature is susceptible to the changes in global circulation and inter-model differences in sub-grid parameterizations.
The day-side mean surface temperature ranges from \SI{261}{\K} in the UM to \SI{268}{\K} in ExoCAM, the latter being the warmest simulation overall (Fig.~\ref{fig:hab1_tseries}b).
ExoCAM-simulated climate is in fact the warmest at almost every height at the substellar point (Fig.~\ref{fig:hab1_substellar}a) and in the troposphere on average (Fig.~\ref{fig:hab1_mm_prof}a).
As Fig.~\ref{fig:hab1_substellar}a shows, LMD-G appears to have an even warmer boundary layer at the substellar point than that in ExoCAM and thus a steep lapse rate of about \SI{\approx 8}{\K\per\km} below \SI{750}{\hecto\pascal}.
The steepest lapse rate, exceeding \SI{\approx 25}{\K\per\km} at the substellar point is produced by the UM, whose boundary layer parameterization is based on a non-local closure and is able to maintain this temperature gradient very close to the surface.
Additionally, the UM is the only THAI GCM that has a non-hydrostatic dynamical core \citep{Turbet21_THAI}, which allows for explicit convection and is another potential reason for the steep lapse rate in this model.
Above the surface layer, all four models have a relatively similar profile of temperature, with the lapse rate fluctuating close to the moist adiabatic lapse rate --- between \SI{7}{\K\per\km} and \SI{4}{\K\per\km} --- throughout most of the troposphere (Fig.~\ref{fig:hab1_substellar}b).

The key distinction between the results from the GCMs is seen at the level of the tropopause (\SI{\approx 60}{\hecto\pascal}), where the lapse rate becomes positive (Fig.~\ref{fig:hab1_substellar}b).
The lowest temperature minimum of \SI{<160}{\K} is predicted by LMD-G, while in the UM it is \SI{\approx 170}{\K}; in ExoCAM and ROCKE-3D the tropopause has the least pronounced cold trap with a temperature minimum of \SI{\approx 180}{\K} (Fig.~\ref{fig:hab1_substellar}a).
Above \SI{\approx 50}{\hecto\pascal}, the temperature begins to rise --- gradually in the ExoCAM, ROCKE-3D and LMD-G simulations and more sharply in the UM --- outlining a stratosphere of TRAPPIST-1e.
The temperature increase in the upper layers is driven by the shortwave absorption, which heats the atmosphere at a rate of up to \SI{15}{\K\per\day} at the substellar point (Fig.~\ref{fig:hab1_substellar}c).
This extreme heating rate is present mostly at the topmost level of the GCMs and so the upper level spacing and the model top height is what partially explains the inter-model differences.
Compared to the dry benchmark case Ben~1 analyzed in \citet{Turbet21_THAI}\footnote{Ben~1 and Ben~2 are dry benchmark cases equivalent to the Hab~1 and Hab~2 cases, but have no moisture. Ben~1 has a 1~bar \ce{N2}-dominated atmosphere with 400~ppm of \ce{CO2}, while Ben~2 has a 1~bar \ce{CO2}-dominated atmosphere.}, shortwave heating rates here are ``bottom-heavy'' meaning that their magnitude is comparable, if not higher, than that at the top of the atmosphere.
This is due to intense absorption of stellar radiation by water vapor concentrated in the lowest half of the atmosphere (\SI{\approx 100}{\hecto\pascal} and below).

The excess of energy received by the day side is redistributed vertically in the troposphere by convection and laterally to the night side by the global atmospheric circulation (see Sec.~\ref{sec:hab1_circ}).
Temperature profiles at the substellar point (Fig.~\ref{fig:hab1_substellar}a) and at its antipode on the night side (not shown) reveal the low spatial variation of the free tropospheric temperature, confirming the weak temperature gradient assumption \citep[similar to earlier studies e.g.][]{Turbet:2016,Boutle2017}.
This can be seen in a relatively low gradient of the tropospheric temperature, whose meridional mean varies from \SI{\approx 230}{\K} on the day side to \SI{\approx 260}{\K} on the night side, with ROCKE-3D having the smallest variation (Fig.~\ref{fig:hab1_mm_prof}a).
The weak variation of tropospheric temperature in ROCKE-3D with longitude is a consequence of a more zonally symmetric global circulation, which is characteristic of synchronously rotating planets in a ``fast rotation'' dynamical regime \citep{Haqq_Misra2018,Sergeev2020}.
This regime is dominated by two extratropical zonal jets (see Sec.~\ref{sec:hab1_circ}) and a stronger equator-to-pole temperature gradient relative to the substellar-antistellar gradient.

Vertically integrated specific humidity, i.e. the water vapor path, is the lowest in ROCKE-3D (up to \SI{12}{\kg\per\m\squared}), whilst the highest amount of water vapor (up to almost \SI{25}{\kg\per\m\squared}) is simulated by ExoCAM and LMD-G  (Fig.~\ref{fig:hab1_mm_prof}b).
In the case of LMD-G, however, moisture is mostly concentrated in the lowest layers close to the substellar point and decreases rapidly with height, making the stratosphere more than an order of magnitude drier compared to the other three models (see last column in Table~\ref{tab:glob_diag}).
As a result, the absorption of shortwave radiation by the water vapor in the stratosphere is the lowest in the LMD-G (Fig.~\ref{fig:hab1_substellar}c).
The low humidity in LMD-G can be explained by the temperature minimum at the tropopause (Fig.~\ref{fig:hab1_substellar}a), which controls the amount of water entering the stratosphere in the convective region.
In addition, the amount of water vapor that reaches the upper atmosphere is strongly affected by the type of a convection parameterization in the GCM.
As mentioned in Sec.~\ref{sec:conv}, LMD-G uses the convective adjustment scheme, which works by adjusting the temperature and humidity on each level towards a pre-defined adiabat.
This scheme does not allow for convective overshooting or entrainment, and so does not affect neutral or stable layers.
ExoCAM, ROCKE-3D and the UM, on the other hand, use the mass-flux parameterizations.
These schemes carry information in the vertical from their original perturbation, modifying the entraining and detraining rates along the way.
If the convective plume is particularly warm and moist when it reaches the tropopause, it can travel a few model levels above it before becoming neutrally buoyant.
During this overshooting, warm and humid air from the plume is heavily detrained into the lower stratosphere.
As a result, the mass-flux schemes tend to deposit more heat and moisture in the upper levels of the model where the adjustment schemes would stop.

\subsubsection{Clouds}  \label{sec:hab1_clouds}
\begin{figure*}
\includegraphics[width=\textwidth]{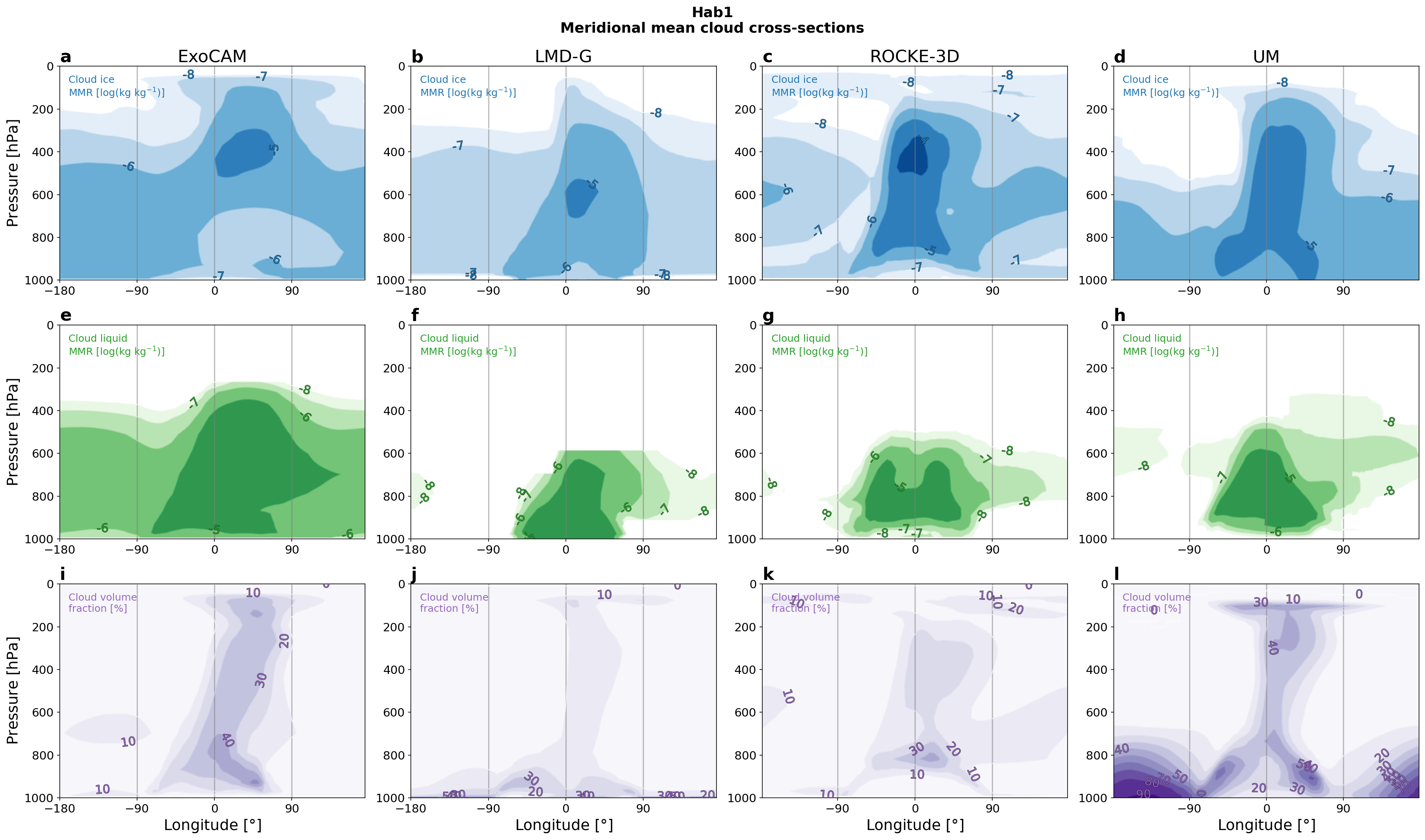}
\caption{Hab~1 results: meridional mean cross-sections of (a--d) mass mixing ratio (MMR) of ice cloud particles ($log[\si{\kg\per\kg}]$), (e--h) MMR of liquid cloud particles ($log[\si{\kg\per\kg}]$), and (i--l) total cloud fraction (\si{\percent}). \label{fig:hab1_cloud_cross}}
\end{figure*}
\begin{figure*}
\includegraphics[width=\textwidth]{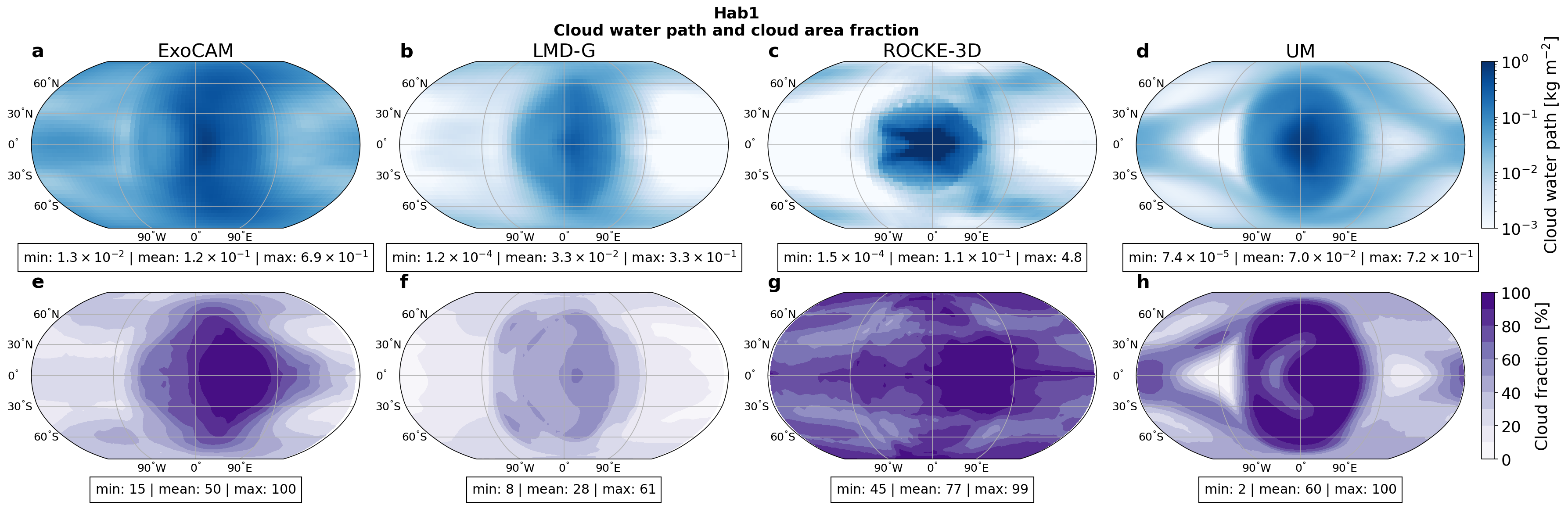}
\caption{Hab~1 results: (a--d) cloud water path (vertically integrated cloud condensate content, \si{\kg\per\m\squared}), and (e--h) cloud fraction (\si{\percent}). \label{fig:hab1_cloud_maps}}
\end{figure*}
Confirming previous studies of cloudy tidally locked planets \citep[e.g.][]{Joshi03,Turbet:2016,Kopparapu2016,Boutle2017,Wolf2017,Turbet:2018aa,Fauchez:2019,Komacek&Abbot2019,Yang2019,Eager_2020}, all our Hab~1 simulations have a deep cloud layer covering the substellar point, demonstrated by the cloud area fraction maximum in Fig.~\ref{fig:hab1_mm_prof}d.
The cloud fraction varies significantly between the four GCMs (Fig.~\ref{fig:hab1_cloud_cross}i--l).
The LMD-G is the least cloudy, having a mean cloud fraction barely above \SI{10}{\percent} throughout most of the atmosphere (Fig.~\ref{fig:hab1_cloud_cross}j).
Its global mean cloud water path, i.e. the vertical integral of cloud condensate, is also the lowest among the models at a value of \SI{3.3e-2}{\kg\per\m\squared} (Fig.~\ref{fig:hab1_cloud_maps}b) which is supported by previous intercomparisons where this model was likewise less cloudy than CAM and other GCMs \citep[e.g.][]{Abbot12,Yang2019}.
ROCKE-3D, on the other hand, has the highest total cloud fraction (\SI{77}{\percent}), presenting TRAPPIST-1e as a planet completely enshrouded in clouds (Fig.~\ref{fig:hab1_cloud_maps}g).
Note that the total cloud fraction in ROCKE-3D appears much higher than in the meridional-mean cross-section Fig.~\ref{fig:hab1_mm_prof}d and \ref{fig:hab1_cloud_cross}c, which may be an artifact of how the model integrates the cloud field vertically.

On the day side, the lowest part of the cloud deck consists of liquid water droplets, while above \SI{\approx 600}{\hecto\pascal} and further away from the substellar point, the clouds are dominated by ice crystals (Fig.~\ref{fig:hab1_cloud_cross}a--h).
The total cloud water path is partitioned differently in the four GCMs: ExoCAM's and LMD-G's clouds are almost entirely comprised of liquid, while ROCKE-3D's clouds are mostly icy, and the UM is roughly half-way between these two extremes.
Despite the high cloud ice content, the longwave CRE in ROCKE-3D is not as high as in ExoCAM, because in the former the cloud water path is concentrated at the substellar point (Fig.~\ref{fig:hab1_cloud_maps}c), exposing the rest of the day side to the loss of thermal radiation.
The eastward tilt of cloudy area in the meridional-mean cross-sections is the consequence of a superrotating equatorial jet and leads to the difference in cloud content at the western and eastern terminators \citep[for more details see Paper III:][]{Fauchez21_THAI}.
Illustrating the process of detrainment of convective cloud condensate, an anvil cloud is visible at the top of the day-side atmosphere in all four GCMs, and is especially pronounced in ExoCAM and the UM (Fig.~\ref{fig:hab1_cloud_cross}i,l).
This anvil is composed of ice crystals (Fig.~\ref{fig:hab1_cloud_cross}a--d) and is the key contributor to the longwave CRE warming the day side, which is indeed the largest in ExoCAM and the UM (Table~\ref{tab:glob_diag}).
LMD-G exhibits a weak cloud anvil and a low-altitude maximum in the icy clouds, which explains the small shortwave and longwave CRE in this model mentioned above.

The largest discrepancy in clouds between our models is on the night side of TRAPPIST-1e.
In the meridional average, the total cloud fraction ranges from \SI{\approx 15}{\percent} (LMD-G) to \SI{\approx 70}{\percent} (ROCKE-3D).
Clouds cover the night side differently: they occur mostly either in the high latitudes (in LMD-G), at the equator (in the UM), or in both regions (in ExoCAM and ROCKE-3D).
They occupy the lowest portion of the troposphere, having the maximum cloud content at \SIrange{1}{3}{\km} above the surface, although the ROCKE-3D case is an outlier that has cloud maximum at \SI{\approx 8}{\km} above the surface (Fig.~\ref{fig:hab1_tseries}k).
Note that even though in ROCKE-3D the cloud content peaks at \SI{\approx 8}{\km}, there are plenty of low-level clouds on the night side too (Fig.~\ref{fig:hab1_cloud_cross}c).
The night side is generally characterized by low-level cloudiness, because convection (especially deep convection) is suppressed, and most night-side clouds are of stratiform type, formed due to the advection and condensation of water vapor from the day side of the planet.
Their formation is largely controlled by the relative humidity, and thus temperature.
Due to a different circulation regime (see Sec.~\ref{sec:hab1_circ}), the temperature distribution in ROCKE-3D has a local minimum in mid-troposphere in high latitudes, just below the eastward jet cores (Fig.~\ref{fig:hab1_zm_u}c).
This results in a local maximum of the relative humidity (not shown), leading to more cloud formation.

From the perspective of the total cloud content and its composition, the ExoCAM case is a distinct outlier, because its cloud water path on the night side is consistently several times higher than that in the other three GCMs (Fig.~\ref{fig:hab1_tseries}g and Fig.~\ref{fig:hab1_mm_prof}c) and because its night-side clouds are dominated by the liquid water (Fig.~\ref{fig:hab1_cloud_cross}e).
Due to the absence of the shortwave radiation on the night side, the water phase of clouds (and thus their optical properties) is less important than their altitude for the net night-side CRE: for ExoCAM, LMD-G, and the UM it is negative, while for ROCKE-3D it is positive but close to zero (Table~\ref{tab:glob_diag}).
The altitude and optical properties of clouds at the terminators affect the transmission spectrum of the atmosphere and so is crucial to simulate correctly in GCMs, because transmission spectroscopy is the primary mode of atmospheric characterization of TRAPPIST-1e.
This is discussed in more detail in Part III \citep{Fauchez21_THAI}.

\subsubsection{Atmospheric circulation} \label{sec:hab1_circ}
\begin{figure*}
\includegraphics[width=\textwidth]{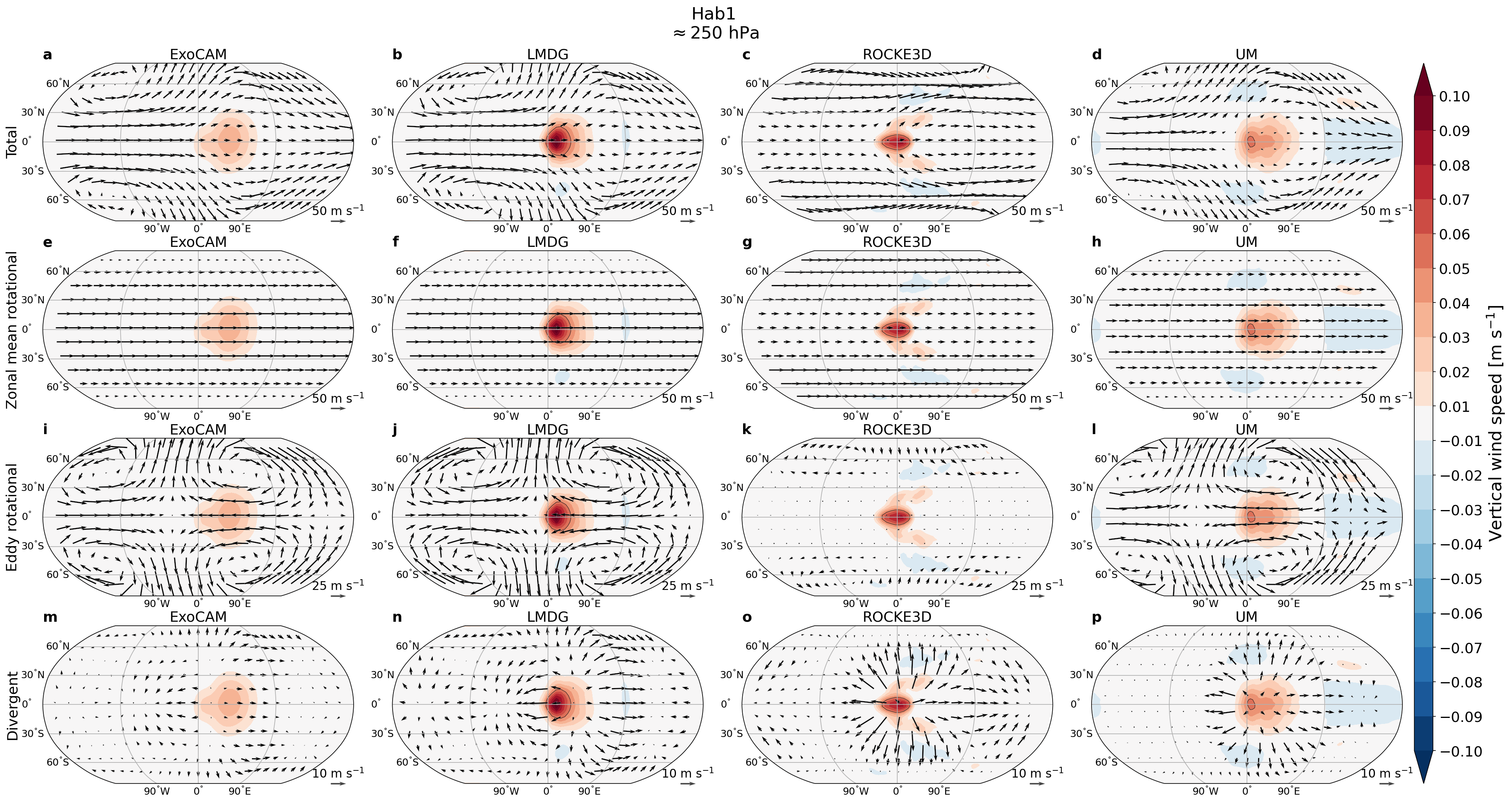}
\caption{Helmholtz decomposition of the horizontal wind at \SI{250}{\hecto\pascal} in the Hab~1 case (quivers): (a--d) total wind, (e--h) zonal mean rotational component, (i--l) eddy rotational component, (m--p) divergent component. Note the different scaling of the eddy rotational and divergent components. Also shown is the upward wind velocity (shading, \si{\m\per\s}) with the \SI{0.05}{\m\per\s} highlighted by a black contour. Note that for ExoCAM, only the pressure velocity ($\omega$, \si{\pascal\per\s}) is available in the output, so the vertical velocity ($w$, \si{\m\per\s}) is approximated as $w=-\omega/\rho g$, where $\rho$ is air density and $g$ is the acceleration due to gravity. \label{fig:hab1_rot_div}}
\end{figure*}
\begin{figure*}
\includegraphics[width=\textwidth]{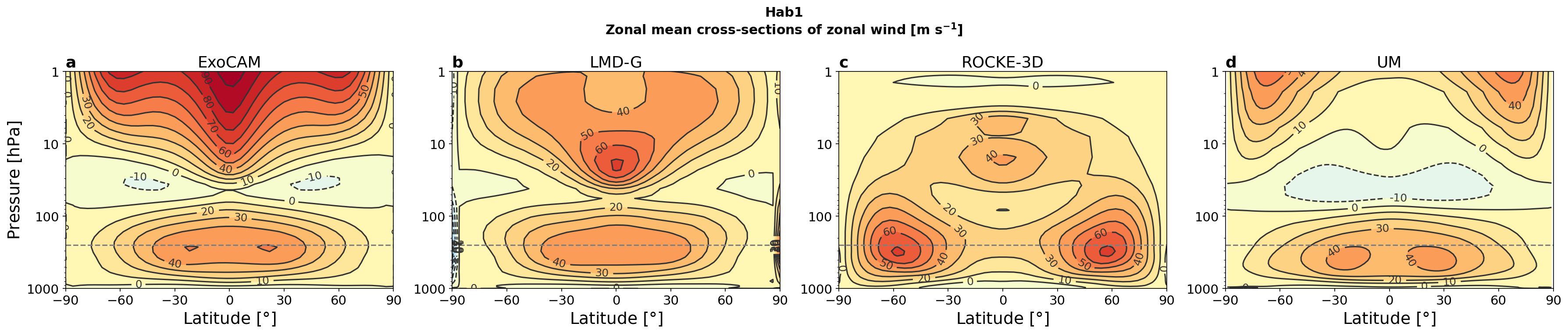}
\caption{Hab~1 results: vertical cross-sections of the zonal mean zonal wind (\si{\m\per\s}). The gray dashed horizontal line marks the \SI{250}{\hecto\pascal} level shown in Fig.~\ref{fig:hab1_rot_div}. \label{fig:hab1_zm_u}}
\end{figure*}

The stationary pattern of the global circulation in the troposphere of tidally locked planets is a combination of several key components.
This is elucidated by Fig.~\ref{fig:hab1_rot_div}, where the Hab~1 wind field at \SI{250}{\hecto\pascal} is split into its rotational and divergent components, using the Helmholtz decomposition methodology of \citet{Hammond21}.
First, on a relatively fast rotating planet such as TRAPPIST-1e, the zonal (eastward) component of the wind forms prograde jets, evident in the zonal mean of the rotational wind field (Fig.~\ref{fig:hab1_rot_div}e--h and \ref{fig:hab1_zm_u}). 
Second, its eddy component (here the deviation from the zonal mean) reveals planetary-scale stationary waves straddling the substellar longitude (Fig.~\ref{fig:hab1_rot_div}i--l).
Third, the divergent wind is relatively weak at this altitude, but clearly shows the outflow branch of the overturning circulation.
At the substellar point, strong divergence aloft is compensated by the convergence in the boundary layer (not shown) and is associated with intense convection (positive upward wind velocity contours in Fig.~\ref{fig:hab1_rot_div}).
Note that even though the magnitude of the divergent wind is small, it is one of the dominant branches of the moist static energy transport between the day and night side on tidally locked planets \citep{Hammond21}.

\begin{figure*}
\includegraphics[width=\textwidth]{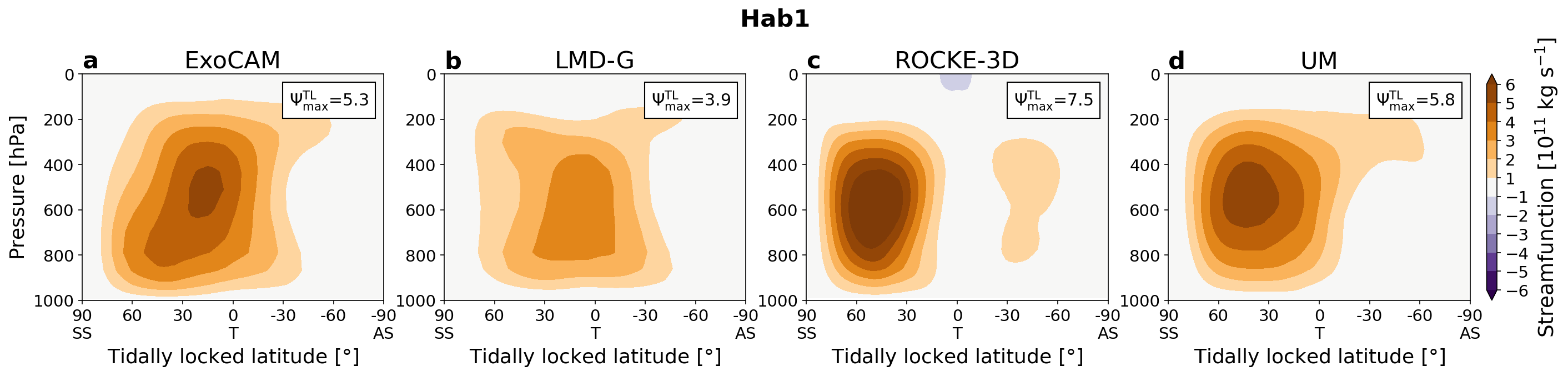}
\caption{Hab~1 results: overturning circulation shown by the tidally locked mass streamfunction $\Psi^{TL}$ (\si{\kg\per\s}). The mass flux is clockwise where $\Psi^{TL}$ is positive and anticlockwise where $\Psi^{TL}$ is negative. The maximum of $\Psi^{TL}$ in each GCM is shown in the upper left corner. Substellar point (SS) is at \ang{90} latitude, antistellar point (AS) is at \ang{-90} latitude. \label{fig:hab1_tl_sf}}
\end{figure*}

The substellar-antistellar overturning circulation is driven by the divergent wind and due to its isotropic structure is best presented in a tidally locked coordinate system \citep{Koll15}.
In this coordinate system, the tidally locked latitude is -90 degrees at the antistellar point, 0 degrees at the North and South poles, and 90 degrees at the substellar point.
The tidally locked longitude is 0 on the line connecting the substellar point, North pole and antistellar point, increasing towards the (original) eastern terminator.
The mass streamfunction $\Psi^{TL}$ in the tidally-locked latitude-pressure plane shows a single circulation cell between the day side and the night side: the air rises at the substellar point, moves toward the antistellar point aloft, subsides on the night side or at the terminators and finally returns to the day side in the lower troposphere (Fig.~\ref{fig:hab1_tl_sf}).

ExoCAM, LMD-G and the UM show a classic wind field pattern in the free atmosphere, consisting of a strong superrotating jet at the equator, stationary gyres in mid-latitudes, and a moderate divergence at the substellar point \citep{Matsuno66_quasi-geostrophic,Gill80_simple}.
Such a pattern emerges in many simulations of moderate climates of tidally locked exoplanets in other 3D GCMs and idealised models \citep[e.g.][]{carone2014connecting,carone2015connecting,carone2016connecting,Haqq_Misra2018,hammond_2020}.
ROCKE-3D, however, produces a different pattern: strong zonal jets in mid-latitudes and a weak superrotation at the equator, reminiscent of the regime of rapidly rotating Earth-like planets (Fig.~\ref{fig:hab1_rot_div}g).
This pattern is associated with a weaker stationary wave response (Fig.~\ref{fig:hab1_rot_div}k), but a stronger substellar divergence than in other GCMs (Fig.~\ref{fig:hab1_rot_div}o).
Correspondingly, the tidally locked mass streamfunction $\Psi^{TL}$ reaches \SI{7.5e11}{\kg\per\s} in ROCKE-3D (Fig.~\ref{fig:hab1_tl_sf}c). 
In the other three models, $\Psi^{TL}=$\SIrange{3.9}{5.8e11}{\kg\per\s} while the overturning cell extends beyond the terminator to the night side (Fig.~\ref{fig:hab1_tl_sf}a,b,d).

The outlier case of ROCKE-3D provides more evidence of the delicate balance in atmospheric circulation of tidally locked but relatively fast rotating planets such as TRAPPIST-1e, as hinted at by earlier studies \citep[e.g.][]{Edson11, Kopparapu2017}.
The global circulation on TRAPPIST-1e can settle on two different regimes: with a dominant eastward jet at the equator (Fig.~\ref{fig:hab1_zm_u}a,b,d) or a pair of mid-latitude eastward jets (Fig.~\ref{fig:hab1_zm_u}c).
This confirms the findings of \citet{Sergeev2020} who demonstrated that this transition is susceptible to change in the parameterization of convection in the UM.
Given the differences in parameterizations between the THAI GCMs, it is unsurprising to capture both circulation regimes in this intercomparison, though it is reassuring to do it with GCMs other than the UM.
A similar dichotomy between circulation regimes can exist for fast-rotating tidally locked planets, assuming both dry and wet atmosphere, as was first shown by \citet{Edson11} using the GENESIS GCM (a fork of the NCAR CESM).
The authors found that for an Earth-like case, an abrupt transition occurs at rotation periods, especially in their wet case (3--4 days), lower than for TRAPPIST-1e (6.1 days, see Table~\ref{tab:planet}).
It is thus possible that due to inter-model differences only ROCKE-3D is susceptible to the circulation change at the rotation period of 6.1 days, while other models would change their circulation if the rotation period is lowered even more.
Another reason for the discrepancy between the studies is that \citet{Edson11} used the stellar spectrum and irradiation of the Sun. However, the study by \citet{carone2014connecting,carone2015connecting,carone2016connecting} used the MIT GCM to explore a broad parameter space of synchronously rotating Earth-like atmospheres; this model configuration included a present-Earth Newtonian relaxation scheme as radiative forcing, which showed this dynamical regime transition to occur at a period of 6--7 days. \citet{carone2014connecting,carone2015connecting,carone2016connecting} attributed this discrepancy with \citet{Edson11} to differences in each model's treatment of the nightside forcing.
Just as in the sensitivity experiments of \citet{Edson11,carone2014connecting,carone2015connecting,carone2016connecting}, and \citet{Sergeev2020}, the two circulation regimes are associated with different cloud patterns, which is indeed seen in ROCKE-3D's cloud fraction map (Fig.~\ref{fig:hab1_cloud_maps}c,g).
The nature of two distinct global circulation regimes for planets like TRAPPIST-1e epitomizes the importance of multi-model studies and is an important avenue of research \citep{Sergeev21_inprep}.

Even more inter-model differences are visible in the upper layers of the atmosphere, as shown in Fig.~\ref{fig:hab1_zm_u}.
At \SI{\approx 50}{\hecto\pascal}, the flow weakens and in two models (ExoCAM and the UM) becomes retrograde; while further above the prograde flow increases again.
This eastward flow takes the shape of two jets near the poles in the UM, while in other models it is concentrated at the equator and is especially strong in ExoCAM (\SI{>100}{\m\per\s}).
The reason why the four GCMs exhibit large differences in the upper atmosphere is the different altitude of the model top, as well as different radiative heating rates (Fig.~\ref{fig:hab1_substellar}).
Another possible factor is the numerical damping at the top of the domain, which is implemented differently in each model \citep{Turbet21_THAI}.


\subsubsection{Time variability} \label{sec:hab1_time_var}
\begin{figure*}
\includegraphics[width=\textwidth]{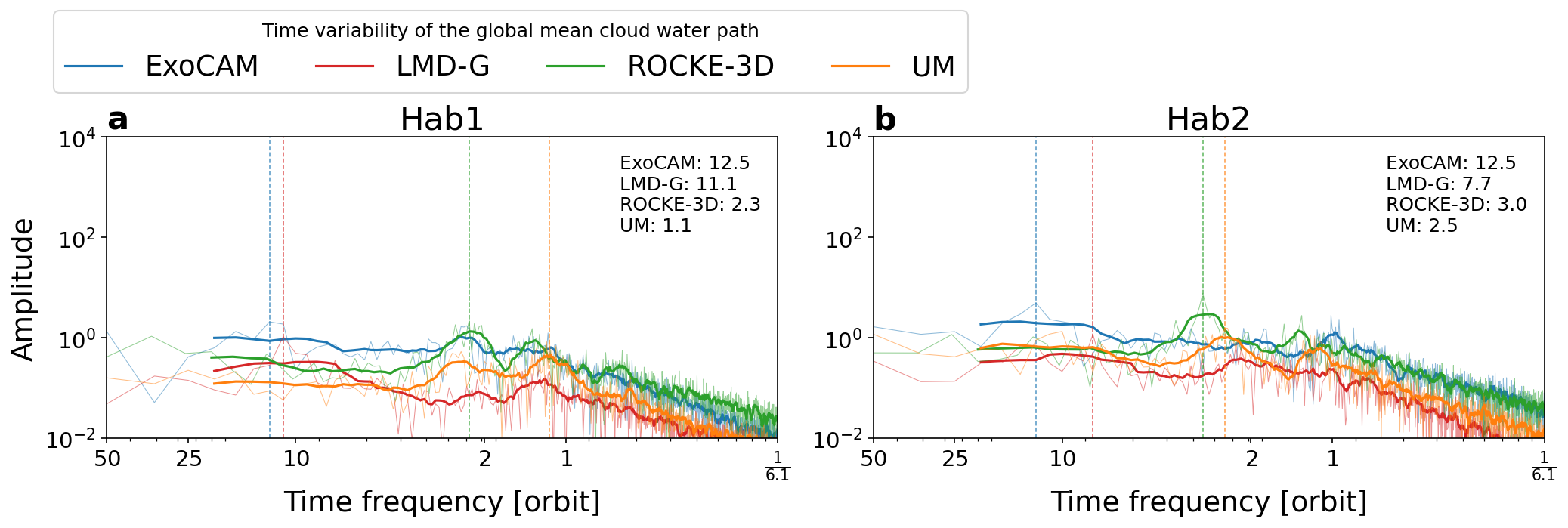}
\caption{Time spectrum of the global mean cloud water path variability. This is the time series shown in Fig.~\ref{fig:hab1_tseries}e and \ref{fig:hab2_tseries}e transformed to a time frequency domain in log-log axes for (a) Hab~1 and (b) Hab~2, respectively. The time frequency units are TRAPPIST-1e orbits (see Table~\ref{tab:planet}). The dashed vertical lines marking the maximum amplitude for each of the models. Periods corresponding to the maximum amplitude are shown in the upper right corner. \label{fig:tfreq_cwp}}
\end{figure*}
Fluctuations around the time mean seen in Fig.~\ref{fig:hab1_tseries} are the manifestation of weather patterns growing and moving in the simulated atmosphere of TRAPPIST-1e.
It is clear that different GCMs have a different amplitude and frequency of the weather variability within the Hab~1 climate.
The largest amplitude of fluctuations is exhibited by ExoCAM, and its principal oscillation period is 12.5 orbits for the global mean cloud content (Fig.~\ref{fig:tfreq_cwp}a).
A secondary peak of variability amplitude can be discerned at about 3 orbits, similar to ROCKE-3D.
LMD-G's variability is overall similar to ExoCAM, albeit at a lower amplitude, while ROCKE-3D and the UM exhibit the lowest period of cloud content fluctuations and a moderate amplitude.
In the UM, the most significant mode of variability are high-frequency fluctuations at 1.1 orbits, though secondary peaks are seen at lower frequency.

The temperature variability in the Hab~1 simulations is also the largest in ExoCAM (Fig.~\ref{fig:hab1_tseries}a).
Its amplitude is the largest on the night side of the planet (Fig.~\ref{fig:hab1_tseries}c) and is partially canceled by the day side variability (Fig.~\ref{fig:hab1_tseries}b).
Preliminary analysis of this variability reveals that it is driven by periodic shifts of the planetary-scale atmospheric waves, resulting in variable heat and water vapor transport to the night side of the planet.
The surface heat budget for the night side is controlled only by the advection of warm and moist air from the day side \citep[e.g.][]{Lewis_2018}, because the ocean heat transport is absent from our simulations.
As the night-side atmosphere becomes warmer and moister, downward fluxes of sensible heat and net LW radiation on the night side increase, raising the surface temperature to \SI{\approx 235}{\K} (Fig.~\ref{fig:hab1_tseries}c).
In the opposite phase of the oscillation, the air becomes drier and colder, allowing the night-side surface temperature to drop to \SI{\approx 215}{\K}.

The period of this global-scale oscillation is about 12.5 TRAPPIST-1e orbits, or approximately 76 Earth days.
This rather long period as well as a preference towards positive zonal wavenumbers (not shown) makes the ExoCAM variability somewhat similar to the Madden-Julian Oscillations (MJO) observed on Earth \citep[see e.g.][]{Madden71_detection,Kiladis09_cckw,Matthews21_dynamical}.
Faithful representation of the MJO remains a challenge even for sophisticated climate and weather models \citep{VallisPenn20_convective}.
In particular, components of the convective parameterization such as entrainment rate and convection triggering affect the initiation and propagation of the MJO \citep[e.g.][]{Holloway13_effects}.
It is thus possible that a combination of moist physics parameterizations and the overall warmer climate in ExoCAM's simulation is able to excite such internal variability, while in other models (e.g. ROCKE-3D and the UM) it is not.
Precisely why this happens in ExoCAM is left for the future, because the amount of additional experiments and outputs required to tackle this question is beyond the scope of the THAI project.

Our time-variability analysis is limited by the 610 Earth-day sample size of the THAI project.
Therefore, the presence of any significant variability on Earth-yearly scales, such as the Quasi-Biennial Oscillation, are missed from our analysis.
On the other hand, the spectra in all four GCMs peaks at 12.5 orbits or less, making the THAI sample able to capture at least some of the important modes of variability.
Time oscillations with shorter periods are also more likely to be observed because of the limitations in telescope time.

\subsection{Global climate in Hab~2 simulations} \label{sec:hab2}
\begin{figure*}
\includegraphics[width=\textwidth]{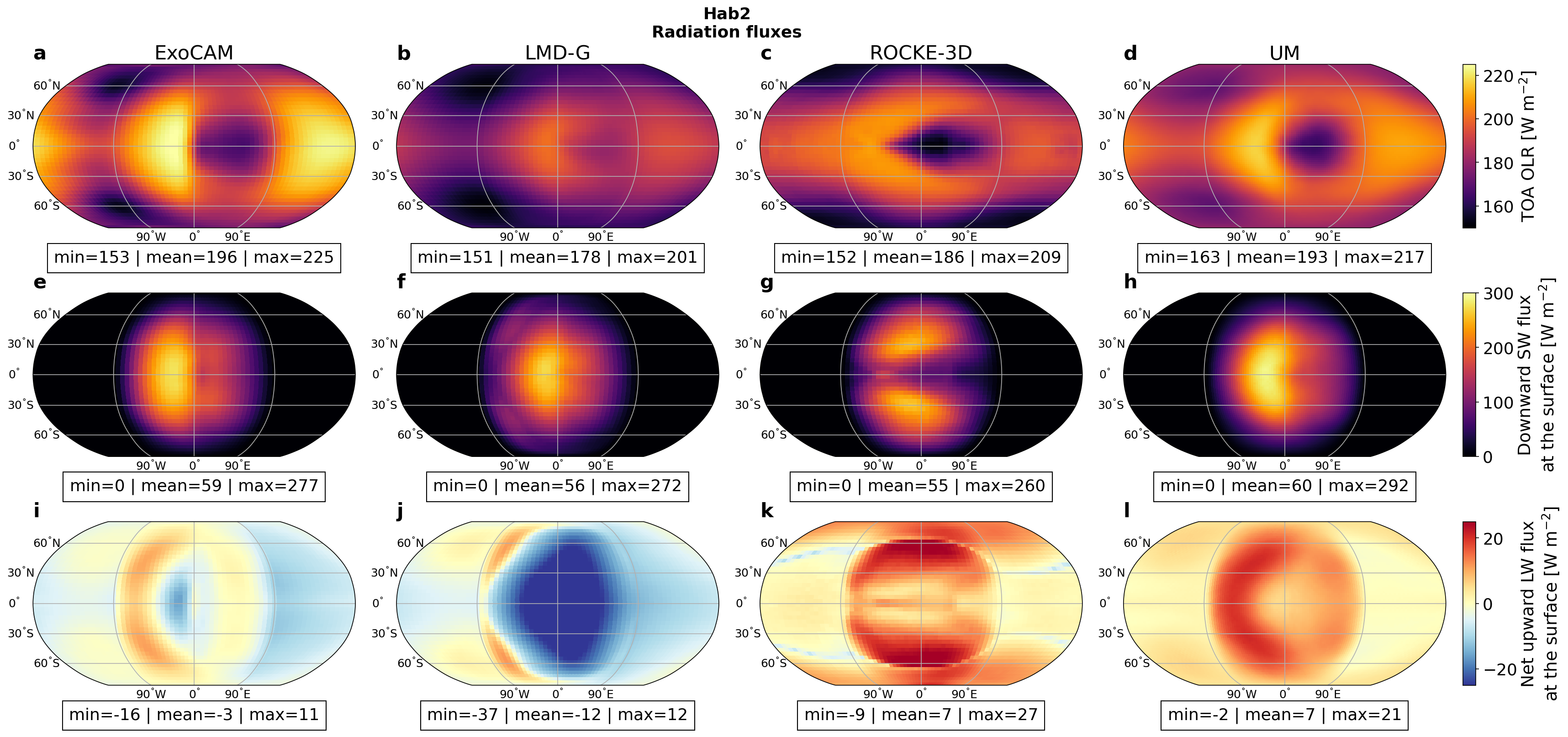}
\caption{Key radiation fluxes in Hab~2 simulations. (a--d) TOA OLR (\si{\watt\per\m\squared}), (e--h) downward shortwave radiation flux at the surface (\si{\watt\per\m\squared}), (i--l) net upward longwave radiation flux at the surface (\si{\watt\per\m\squared}). \label{fig:hab2_rad_flux}}
\end{figure*}
\begin{figure*}
\includegraphics[width=\textwidth]{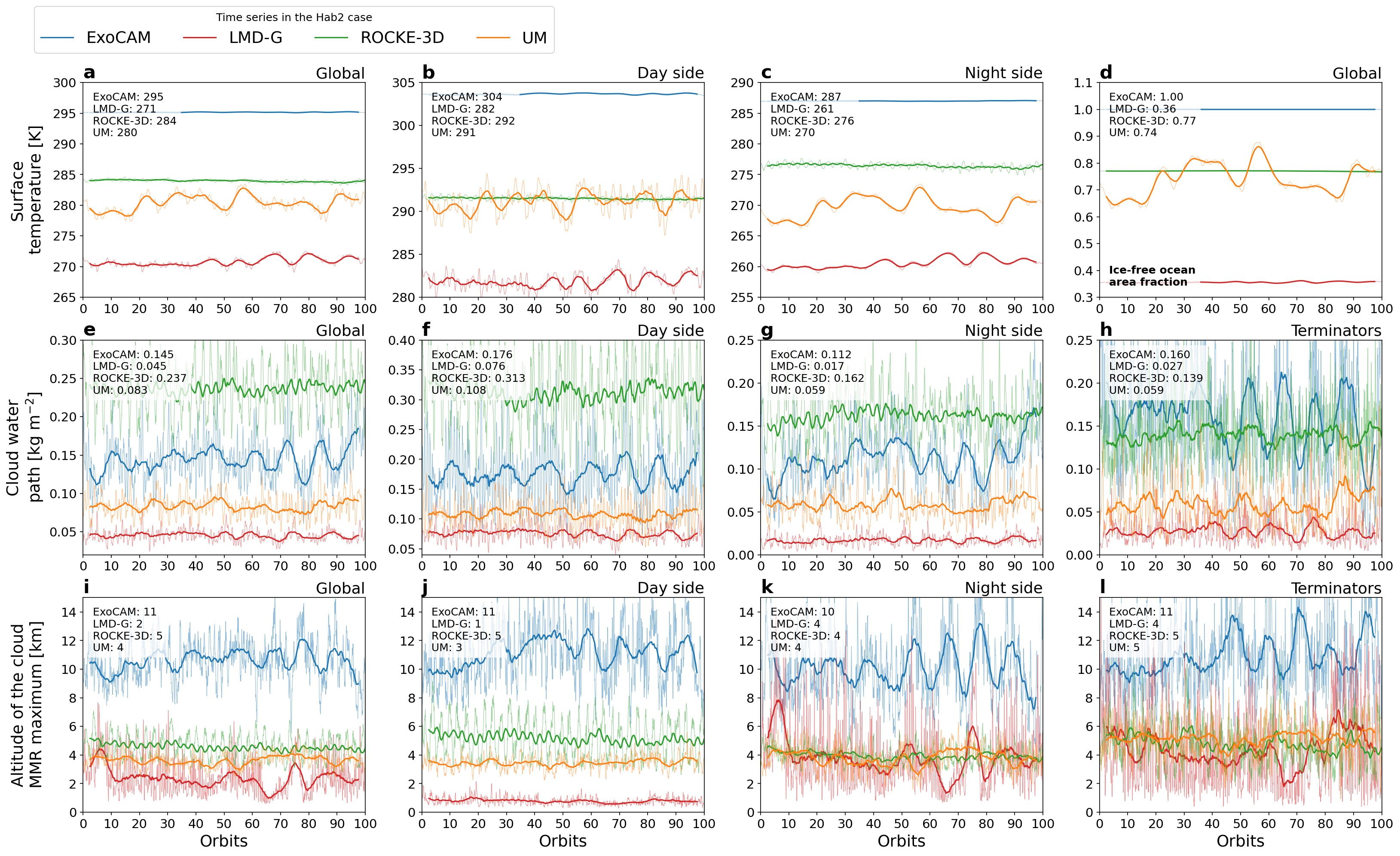}
\caption{Hab~2 results: time variability of (a--c) surface temperature (\si{\kelvin}), (d) ice-free ocean area fraction, (e--h) cloud water path (\si{\kg\per\m\squared}) and (i--l) the altitude of the cloud content maximum (\si{\km}). Each column shows (a, d, e, i) global mean, (b, f, j) day-side mean, (c, g, k) night-side mean and (h, l) terminator mean. Thin lines show the raw data, thick lines show a 5-orbit rolling mean.
\label{fig:hab2_tseries}}
\end{figure*}
\begin{figure*}
    \centering
    \includegraphics[width=\textwidth]{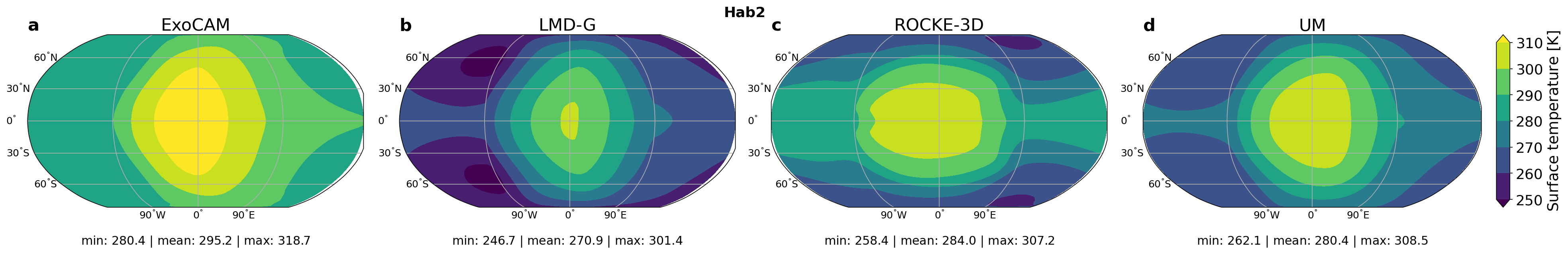}
    \caption{Hab~2 results: time mean surface temperature (\si{\K}) in (a) ExoCAM, (b) LMD-G, (c) ROCKE-3D, and (d) the UM.
    \label{fig:hab2_t_sfc_map}}
\end{figure*}
\begin{figure*}
\includegraphics[width=\textwidth]{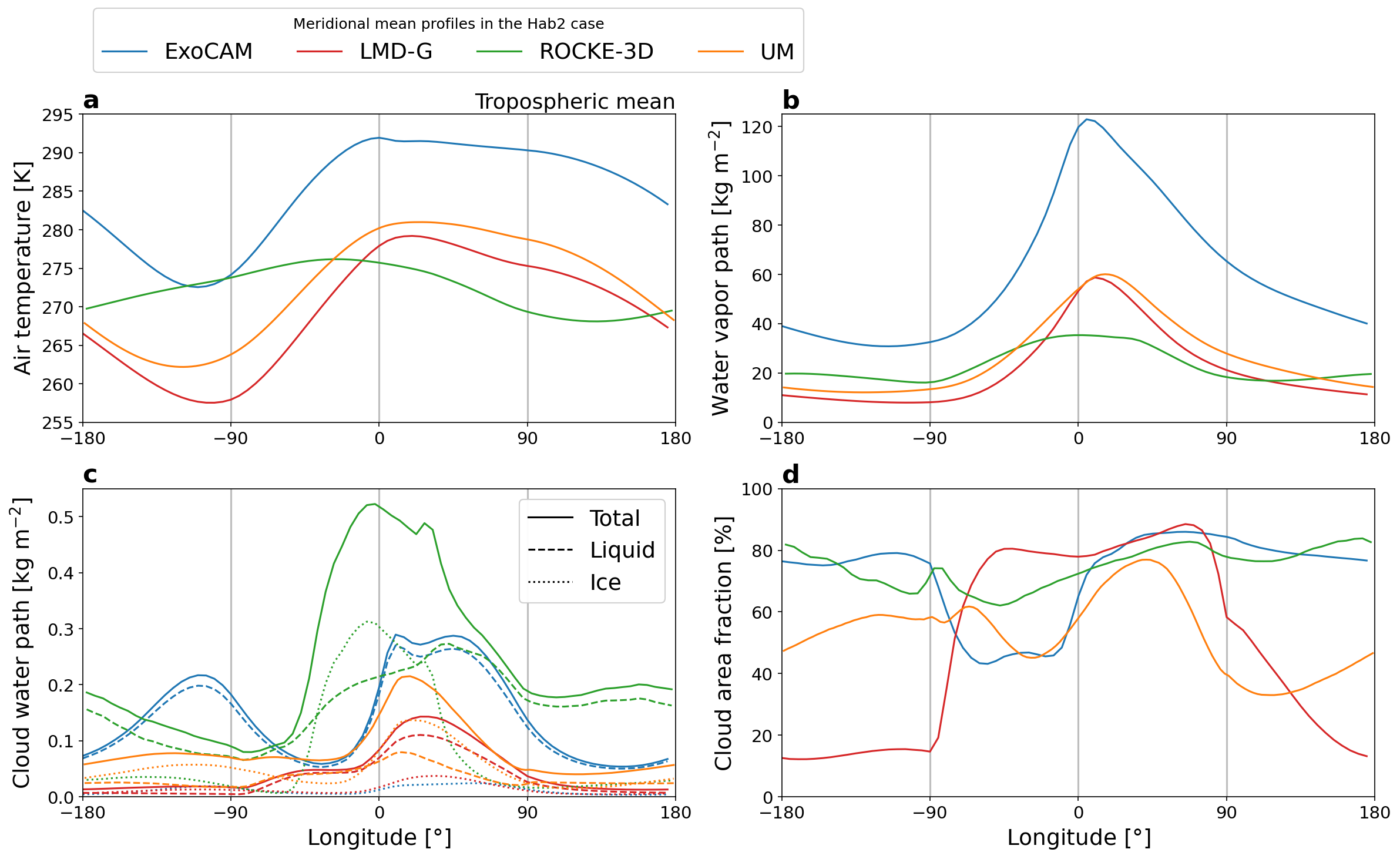}
\caption{Hab~2 results: meridional mean profiles of (a) the mass-weighted tropospheric mean of air temperature (\si{\kelvin}), (b) water vapor path (\si{\kg\per\m\squared}), (c) cloud water path (\si{\kg\per\m\squared}): total (solid lines), liquid (dashed lines), ice (dotted lines), and (d) cloud area fraction (\si{\percent}). Here, the troposphere's boundary is defined by the lapse rate decreasing below \SI{2}{\K\per\km} in the upper atmosphere. Note the different y-axis limits compared to Fig.~\ref{fig:hab1_mm_prof}. \label{fig:hab2_mm_prof}}
\end{figure*}
\begin{figure*}
\includegraphics[width=\textwidth]{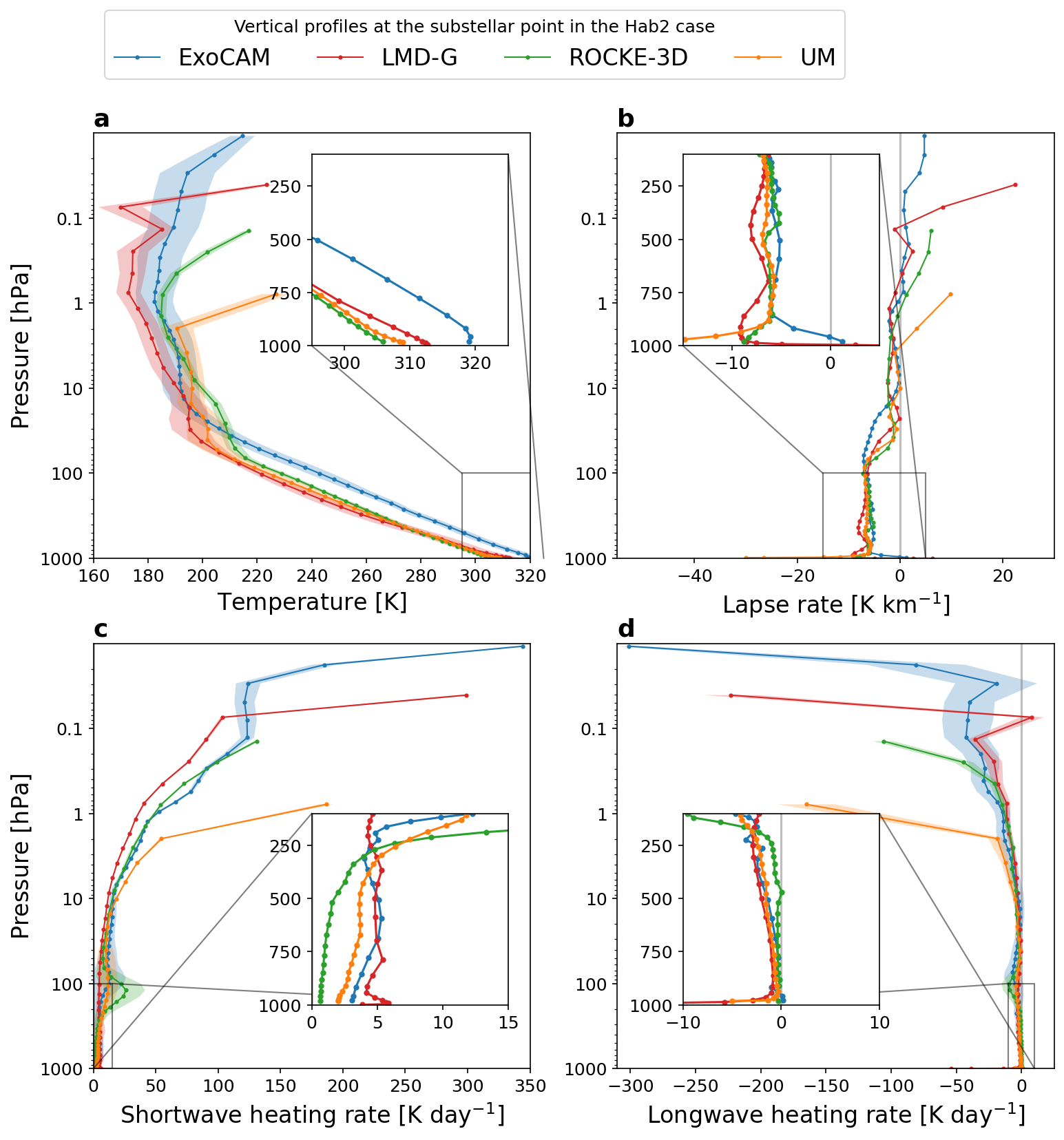}
\caption{Hab~2 results: substellar vertical profiles of (a) air temperature (\si{\kelvin}), (b) lapse rate (\si{\kelvin\per\km}), (c) SW heating rate (\si{\kelvin\per\day}), and (d) LW heating rate (\si{\kelvin\per\day}).
Time mean values are shown by solid lines, the 1-$\sigma$ deviation is shown by shading.
\label{fig:hab2_substellar}}
\end{figure*}
\begin{figure*}
\includegraphics[width=\textwidth]{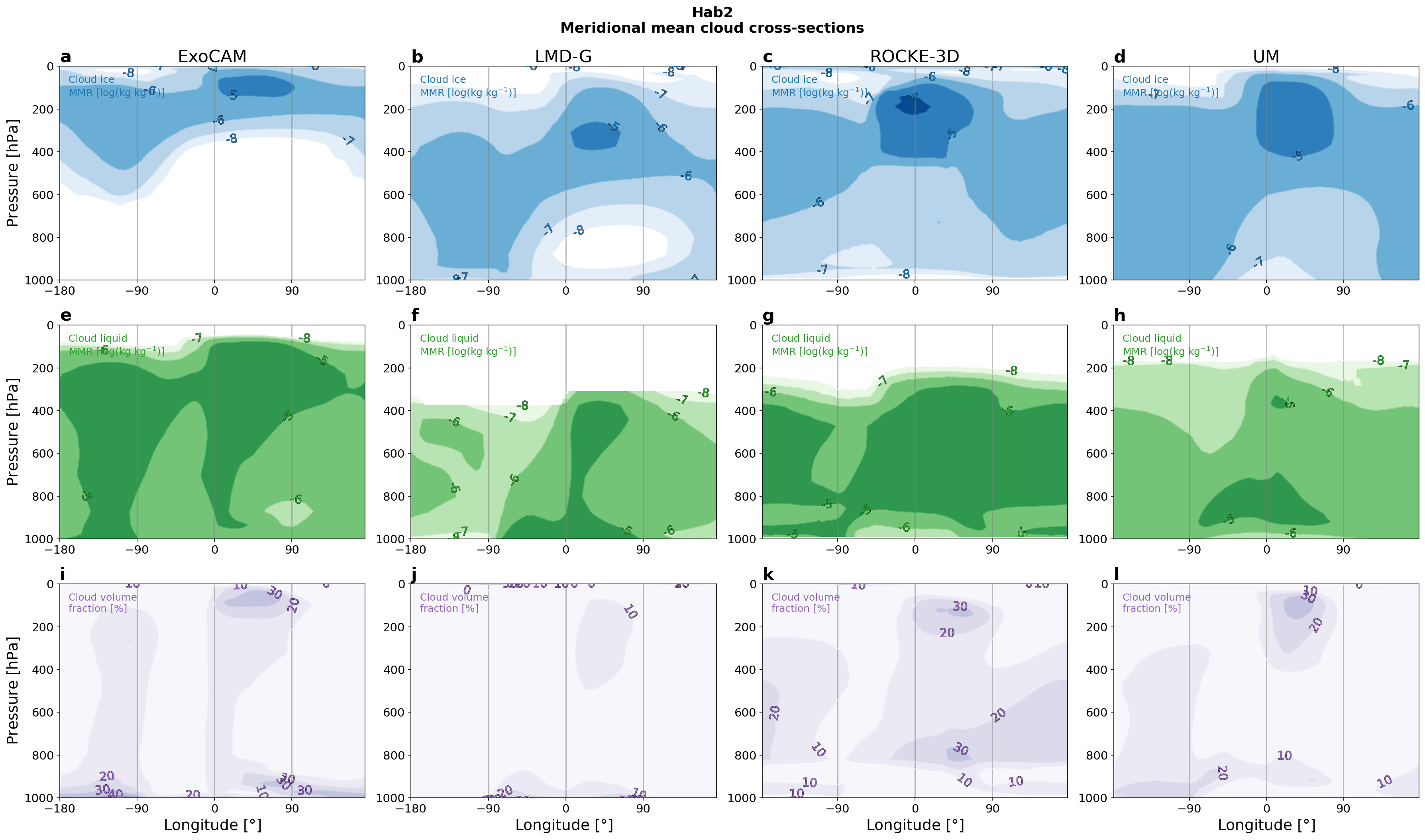}
\caption{Hab~2 results: meridional mean cross-sections of (a--d) mass mixing ratio (MMR) of ice cloud particles ($log[\si{\kg\per\kg}]$), (e--h) MMR of liquid cloud particles ($log[\si{\kg\per\kg}]$), and (i--l) total cloud fraction (\si{\percent}).
\label{fig:hab2_cloud_cross}}
\end{figure*}
\begin{figure*}
\includegraphics[width=\textwidth]{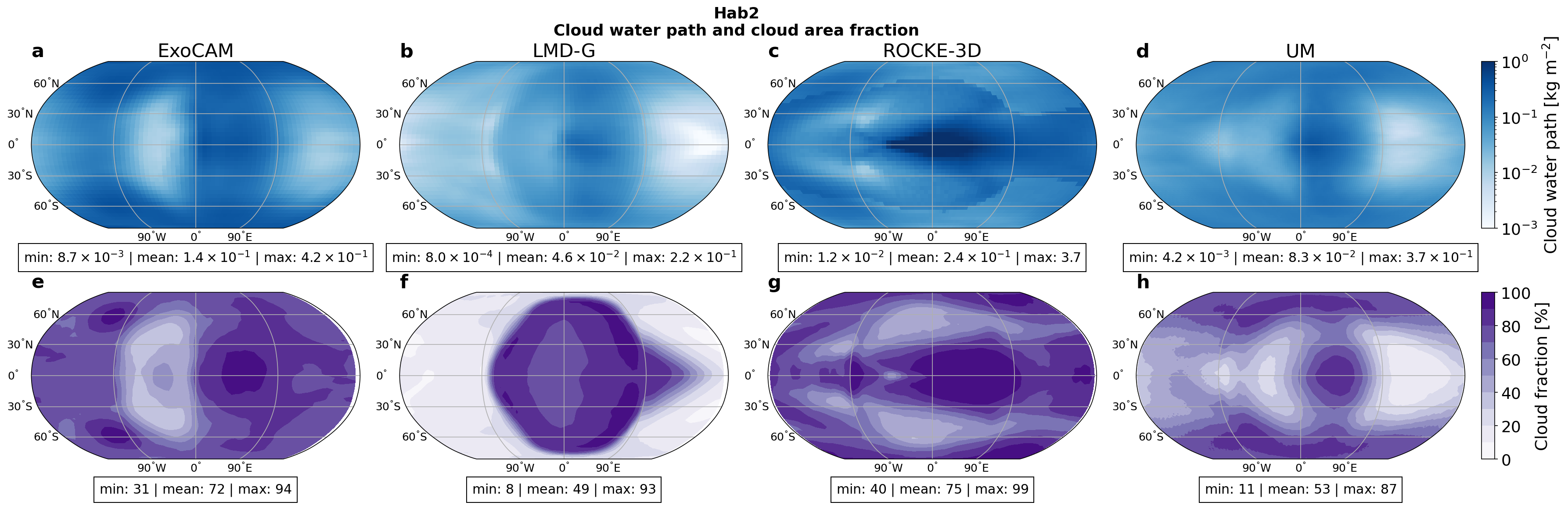}
\caption{Hab~2 results: (a--d) cloud water path (vertically integrated cloud condensate content, \si{\kg\per\m\squared}), and (e-h) cloud fraction (\si{\percent}). \label{fig:hab2_cloud_maps}}
\end{figure*}
\begin{figure*}
\includegraphics[width=\textwidth]{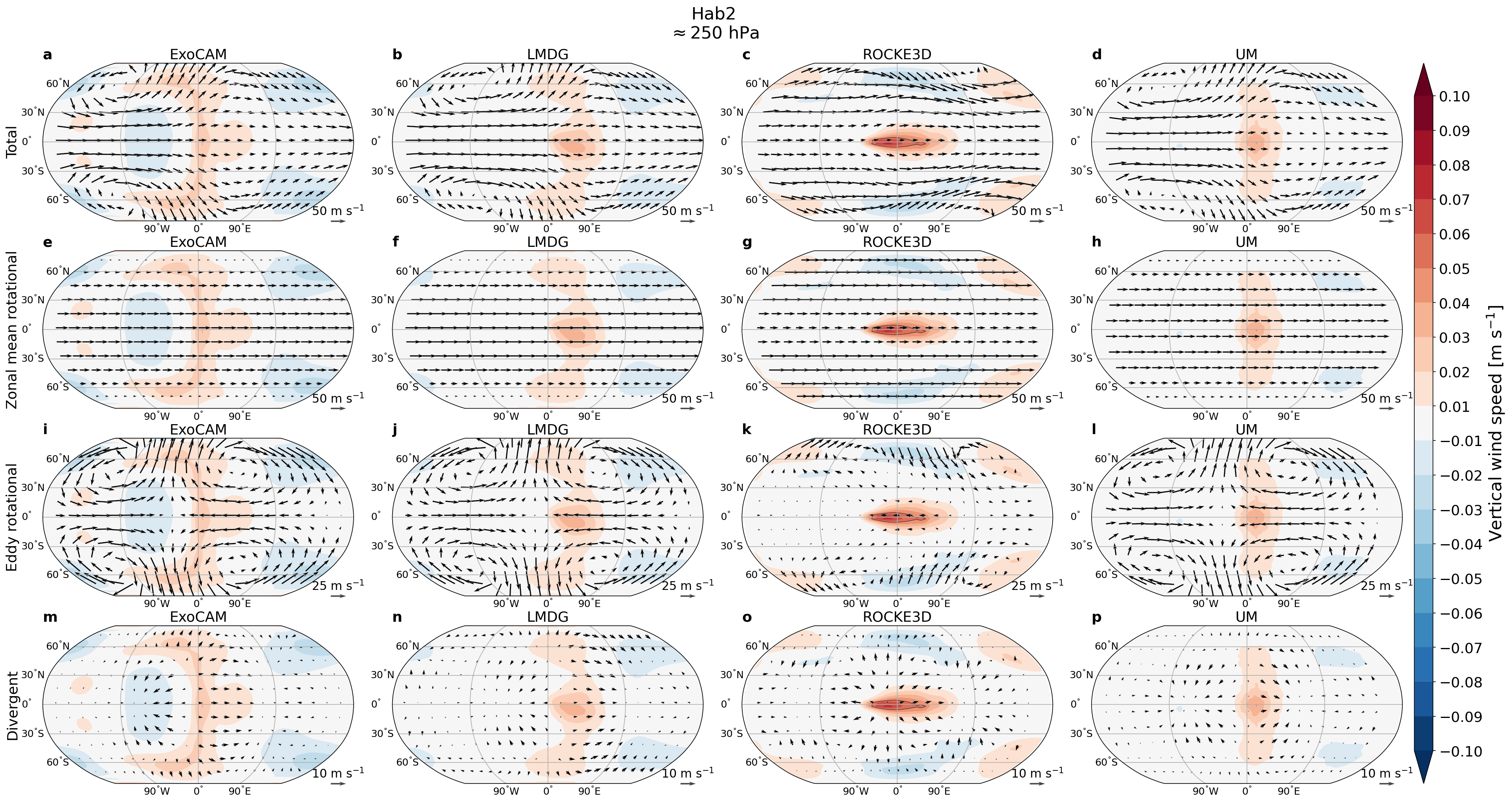}
\caption{As in Fig.~\ref{fig:hab1_rot_div}, but for Hab~2. \label{fig:hab2_rot_div}}
\end{figure*}
\begin{figure*}
\includegraphics[width=\textwidth]{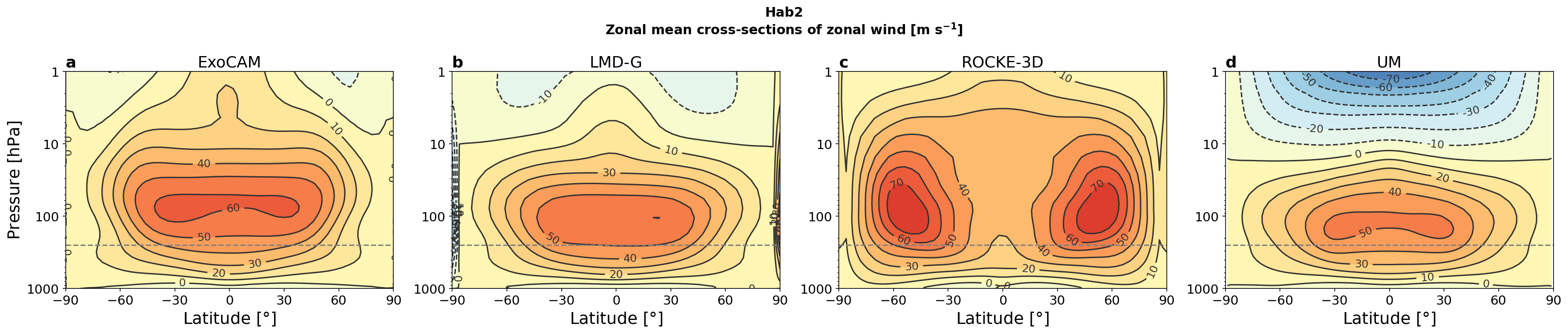}
\caption{Hab~2 results: vertical cross-sections of the zonal mean zonal wind (\si{\m\per\s}). The gray dashed horizontal line marks the \SI{250}{\hecto\pascal} level shown in Fig.~\ref{fig:hab2_rot_div}. \label{fig:hab2_zm_u}}
\end{figure*}
\begin{figure*}
\includegraphics[width=\textwidth]{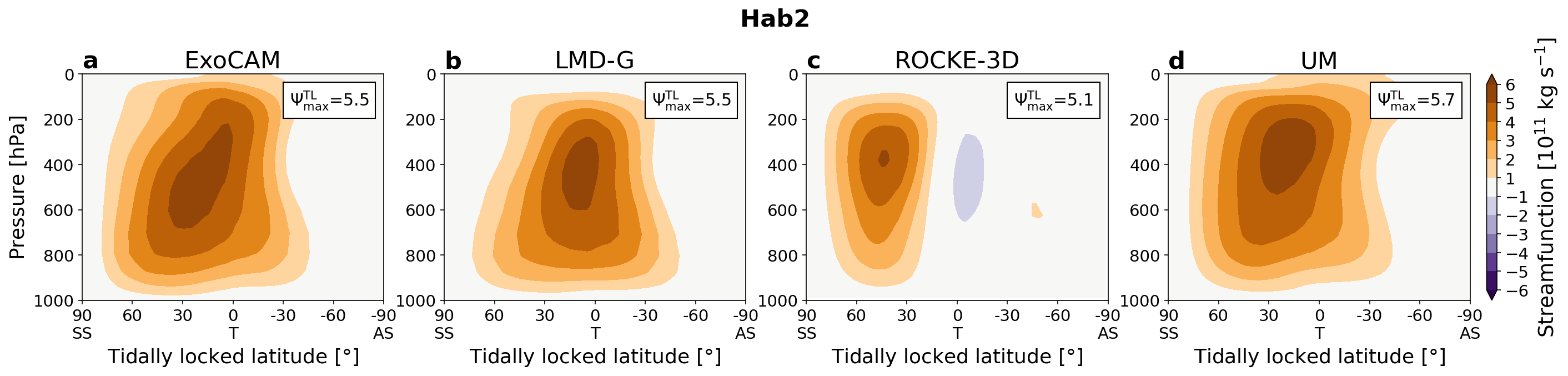}
\caption{As in Fig.~\ref{fig:hab1_tl_sf}, but for Hab~2. \label{fig:hab2_tl_sf}}
\end{figure*}

\subsubsection{Radiation fluxes and thermodynamic profiles} \label{sec:hab2_thermo}
At the top of the \ce{CO2}-dominated atmosphere of Hab~2, the absorbed shortwave radiation is larger than that for Hab~1 by \SIrange{2}{32}{\watt\per\m\squared}, but the overall pattern is the same (not shown).
This increase is mostly due to stronger absorption by \ce{CO2} and \ce{H2O} and, in all models except LMD-G, due to a weaker CRE\textsubscript{SW} (Table~\ref{tab:glob_diag}).
In LMD-G, CRE\textsubscript{SW} reaches \SI{-31.8}{\watt\per\m\squared} and is the second largest after that in ROCKE-3D (\SI{-35.0}{\watt\per\m\squared}).
Accordingly, the Bond albedo $\alpha_\text{p}$ of TRAPPIST-1e is the highest in these two GCMs compared to ExoCAM and the UM, though it is still lower than that in Hab~1 (Table~\ref{tab:glob_diag}).

The longwave part of CRE in Hab~2 is somewhat smaller compared to that in Hab~1, but is likewise the lowest in LMD-G and the highest in ExoCAM (Table~\ref{tab:glob_diag}).
The spatial pattern of the TOA OLR is broadly similar to that in Hab~1 (Fig.~\ref{fig:hab2_rad_flux}a--d).
However, the contrast between the day side and the night side is reduced dramatically.
What ExoCAM illustrates in particular, is that the OLR has two sharp peaks in its longitudinal distribution, thus having an imprint in the emission spectra.
This would also be apparent in broadband spectra for which the minimum would be shifted eastward and the maximum would be closer to \ang{90} (not shown).

The net effect of clouds on the TOA radiation fluxes is still negative, though much weaker relative to that in Hab~1, especially in the cases of ExoCAM and the UM (Table~\ref{tab:glob_diag}).
The exception here is LMD-G, which due to its stronger shortwave part has the net CRE of \SI{-28.6}{\watt\per\m\squared}.
Thus, in three out of four models clouds play a lesser role in controlling energy fluxes in the climate of a \ce{CO2}-dominated atmosphere (Hab~2) than that in a \ce{N2}-dominated atmosphere (Hab~1); in LMD-G though the trend is the opposite.

The distribution of shortwave radiation at the surface of the planet reaches its maximum to the west of the substellar point (Fig.~\ref{fig:hab2_rad_flux}e--h).
This flux is the lowest in ROCKE-3D and the highest in the UM, with their respective distribution nevertheless being similar to those in Hab~1.
The average incident shortwave radiation is similar in all four models, between \SIrange{55}{60}{\watt\per\m\squared} and almost twice as low as in Hab~1 due to larger absorption of radiation by the atmosphere.
Note that the bulk of the shortwave radiation is absorbed by the oceanic surface with a relatively low albedo of 0.06, because most of the day side is above the freezing point of water, which is the threshold for the albedo change in the THAI protocol (Fig.~\ref{fig:hab2_tseries}b).
While the net downward surface shortwave radiation flux is positive, its longwave counterpart is positive only in the Hab~2 cases of ROCKE-3D and UM.
In ExoCAM and especially in LMD-G, the net upward longwave flux at the surface is negative at almost all grid points (Fig.~\ref{fig:hab2_rad_flux}j), implying that the atmosphere is radiating heat to the surface --- in stark contrast to the Hab~1 simulations (Fig.~\ref{fig:hab1_rad_flux}i--l) and the Hab~2 simulations of ROCKE-3D and the UM (Fig.~\ref{fig:hab2_rad_flux}k,l).
This is likely due to the differences in the gaseous absorption in the radiative transfer schemes.
ROCKE-3D and the UM employ the same scheme and the same set of spectral files (see Sec.~\ref{sec:rt}), thus simulating overall similar (weaker) atmospheric absorption (Fig.~\ref{fig:hab2_substellar}c), while ExoCAM and LMD-G use schemes that tend to overestimate shortwave absorption, especially by \ce{H2O} in the case of ExoCAM.

At the substellar point, the surface temperature exceeds \SI{300}{\K} in all four models, but the range of its maximum values is wider than that in the Hab~1 case: from \SI{301}{\K} in LMD-G to \SI{319}{\K} in ExoCAM (Fig.~\ref{fig:hab2_t_sfc_map}).
The ExoCAM simulation is the warmest throughout all of the troposphere at the substellar point\footnote{The ExoCAM THAI simulations use an old version of the radiative transfer code, which overestimates the shortwave absorption by \ce{H2O} and the longwave absorption by \ce{CO2} for dense atmospheres. The new THAI simulations were not ready to be included in the intercomparison in time, but preliminary results show that with the updated \ce{CO2} radiative transfer, ExoCAM gets a mean surface temperature of \SI{\approx 282}{\K} and as such is no longer an outlier in Hab~2.}, before being overtaken above by ROCKE-3D and the UM (Fig.~\ref{fig:hab2_substellar}a).
The transition between the troposphere and stratosphere is located slightly higher and is less pronounced in Hab~2 than in Hab~1.
In the upper atmosphere, the heat balance is primarily between the absorption of stellar radiation and emission of planetary radiation, both of which are an order of magnitude larger than those in Hab~1 (Fig.~\ref{fig:hab2_substellar}c,d).
Due to strong absorption of stellar flux by \ce{CO2} in the Hab~2 case, overwhelming the absorption by water vapor in the troposphere, the radiative heating profile is larger at higher altitudes, just as in the benchmark simulations \citep{Turbet21_THAI}.
Compared to the benchmark simulations, both Hab~1 and Hab~2 cases see larger differences in the radiative heating profiles between the GCMs (even those employing the same RT code, as is the case for ROCKE-3D and the UM), which is due to the radiative effects of water vapor and clouds.

Due to heavily enhanced greenhouse effect of the \ce{CO2} atmosphere (the $G$ parameter in Table~\ref{tab:glob_diag}), the global surface temperature of the Hab~2 cases are \SIrange{40}{50}{\K} higher than those of Hab~1 (Fig.~\ref{fig:hab2_t_sfc_map}).
This is mostly due to the significant warming of the night side of the planet (Fig.~\ref{fig:hab2_tseries}c).
The inter-model spread in predicted surface temperature is larger than that in the Hab~1 simulations, slightly more on the day side and slightly less on the night side (cf. Fig.~\ref{fig:hab1_tseries}b,c and Fig.~\ref{fig:hab2_tseries}b,c).
This spread reduces somewhat when the whole troposphere is concerned, which is substantially warmer than that for the Hab~1 case (Fig.~\ref{fig:hab2_mm_prof}a).
ExoCAM's troposphere is warmer than those of the other models and thus contains a markedly larger amount of water vapor (Fig.~\ref{fig:hab2_mm_prof}b).
The water vapor path is an order of magnitude larger in Hab~2 than in Hab~1 due to the Clausius-Clapeyron law.

\subsubsection{Clouds} \label{sec:hab2_clouds}
Despite the climate being much warmer, the Hab~2 simulations bear many similarities with those of Hab~1 in terms of cloud patterns.
The bulk of cloud water is on the day side of the planet in all four GCMs (Fig.~\ref{fig:hab2_mm_prof}c), having the average day side cloud water path of \SIrange{0.1}{0.3}{\kg\per\m\squared} (Fig.~\ref{fig:hab2_tseries}f,g).
However, vertically integrated cloud diagnostics, such as the cloud water path or the cloud area fraction, show that TRAPPIST-1e is cloudier in the Hab~2 case than in the Hab~1 case, especially on the night side of the planet  (Fig.~\ref{fig:hab2_mm_prof}c,d and \ref{fig:hab2_tseries}g).

For Hab~2 simulations the clouds are more extended in the vertical, which is visible in the meridional mean cross-sections (Fig.~\ref{fig:hab2_cloud_cross}).
In fact, non-negligible cloud condensate extends to the top of the model domain in two GCMs: ExoCAM and LMD-G.
From an observational perspective, simulated transmission spectra are affected differently when the upper atmosphere is cloudy, so using ExoCAM or LMD-G may contradict the interpretations obtained from ROCKE-3D or the UM \citep[see Paper III:][]{Fauchez21_THAI}.
The global mean altitude of the cloud content maximum is also higher in the Hab~2 case relative to that in Hab~1 in three out of four models: ExoCAM (most pronounced), LMD-G, and the UM (Fig.~\ref{fig:hab2_tseries}i).
While ROCKE-3D predicts the most amount of clouds among the models (Fig.~\ref{fig:hab2_tseries}e, \ref{fig:hab2_cloud_maps}c,g), its clouds are closer to the surface, especially liquid clouds on the night side (Fig.~\ref{fig:hab2_cloud_cross}g).
The second cloudiest simulation is by ExoCAM, where the cloud content (mostly liquid droplets) reaches its maximum at the highest altitude (\SI{11}{\km}) among the four GCMs (Fig.~\ref{fig:hab2_tseries}i).
Interestingly, the minimum of cloudiness in ExoCAM is not on the night-side as in other models, but in the western half of the day hemisphere (e.g. Fig.~\ref{fig:hab2_cloud_cross}a,e,i), which may have a detectable imprint in the reflected star light \citep{Wolf2019}.

\subsubsection{Atmospheric circulation} \label{sec:hab2_circ}
The tropospheric wind structure in all four Hab~2 simulations is similar to that in their respective Hab~1 simulations (cf. Fig.~\ref{fig:hab1_rot_div}a--d and Fig.~\ref{fig:hab2_rot_div}a--d).
ExoCAM, LMD-G and the UM develop the equatorial jet pattern, while ROCKE-3D develops two extratropical jets (Fig.~\ref{fig:hab2_zm_u}).
Accordingly, the day-side vertical velocity maximum is more pronounced at the equator in the ROCKE-3D simulation and spread out across latitudes in the other three simulations (color shading in Fig.~\ref{fig:hab2_rot_div}).

Fig.~\ref{fig:hab2_tl_sf} shows that the strength of the day-night circulation does not change significantly compared to that of the Hab~1 case.
The mass streamfunction $\Psi^{TL}$ in each GCM has broadly the same shape in both the Hab~1 and Hab~2 experiments.
Namely, $\Psi^{TL}$ in LMD-G is bottom-heavy, showing a strong flow towards the substellar point in the boundary layer, while in the UM it is top-heavy, showing strong divergence in the upper atmosphere.
As is the case for Hab~1, the ROCKE-3D's overturning cell in the Hab~2 simulations is confined to the day side, while in the other three GCMs it extends to the night side \citep[as in e.g.][]{ZhangYang20,Hammond21}.

In the troposphere, the differences in the mean eastward wind speed between the Hab~1 and Hab~2 cases are rather small, with the maximum value staying within \SIrange{50}{70}{\m\per\s} (Fig.~\ref{fig:hab2_zm_u}).
The stratospheric circulation, however, differs more noticeably: in all four models the uppermost levels do not develop an additional region of superrotation (not shown).
Instead, they are dominated by a westward flow, which is especially strong in the UM (Fig.~\ref{fig:hab2_zm_u}d).
As discussed in more detail in Part I \citep{Turbet21_THAI}, the source of inter-model differences near the top lid is due to the differences in clear-sky radiative transfer, location and spacing of vertical levels, and numerical damping (sponge layers).
For example, ROCKE-3D has relatively high numerical damping coefficients to increase model stability, which results in a low wind speed near the top boundary compared to that in the other three GCMs (Fig.~\ref{fig:hab2_zm_u}c).
The impact on the stratospheric circulation by the numerical drag has been discussed in \citet{Carone2018}, but this question would further benefit from a dedicated GCM intercomparison.
Thus, the overall characteristic of the time-mean zonal flow for the Hab~2 simulations is that a single structure of the equatorial (ExoCAM, LMD-G, the UM) or extratropical (ROCKE-3D) jet extends throughout most of the lower atmosphere.

\subsubsection{Time variability}
\label{sec:hab2_time_var}
Compared to Hab~1 case, the warmer climate of the Hab~2 case sees an increase of the vapor and condensed water at the terminators (Fig.~\ref{fig:hab2_tseries}), which consequently leads to an increased amplitude of the cloud content variability (Fig.~\ref{fig:tfreq_cwp}b).
The periodicity of these fluctuations, however, stays roughly the same for each of the GCMs, with relatively longer periods in ExoCAM and LMD-G and shorter periods in ROCKE-3D and the UM.
This is also true for the global surface temperature fluctuations (spectra not shown, but compare Fig.~\ref{fig:hab1_tseries}a to Fig.~\ref{fig:hab2_tseries}a), though for LMD-G and the UM they exhibit longer periods than those for the cloud content.
Unlike the cloud variability, the amplitude of temperature variability in the ExoCAM simulation of Hab~2 drops by more than an order of magnitude compared to that of Hab~1, which was noted in the preliminary THAI results \citep{Fauchez2020THAI}.
However, it is not the case for the other GCMs: the amplitude of temperature fluctuations increases in the UM and LMD-G and stays the same in ROCKE-3D (not shown).

\section{Synthesis and discussion} \label{sec:synthesis}
\subsection{Inter-model similarities and differences}
To a large degree, all four THAI GCMs predict temperatures within \SIrange{20}{25}{\K} of each other in both the Hab~1 and Hab~2 cases.
On the largest scale, the atmosphere stays habitable and sustains a stable water cycle without entering a moist greenhouse state or an atmospheric collapse due to the \ce{CO2} condensation (see Sec.~\ref{sec:habitability}).
While the inter-model spread in the global mean surface temperature is non-negligible, it is smaller in the Hab~1 case (\SI{14}{\K}) and larger in Hab~2 (\SI{24}{\K}, see Table~\ref{tab:glob_diag}).
These differences are two and four times larger than those found in the Ben~1 and Ben~2 simulations, respectively \citep{Turbet21_THAI}, demonstrating the amplification of inter-model spread by moist physics in GCMs.
Our results suggest that while in the dry climates the inter-model spread is similar \citep{Turbet21_THAI}, in moist climates it is larger for a warmer case --- at least for the four THAI GCMs.
The inter-model spread is smaller on the day side (especially in the Hab~1 case) and larger on the night side of the planet.
Interestingly, the largest temperature difference in the Hab~2 simulations is between ExoCAM and LMD-G.
These two models were at the extremes in a previous exoplanet GCM intercomparison \citep{Yang2019}, although LMD-G was warmer than ExoCAM, whereas in our study it is the opposite.

Global mean energy fluxes at the top of the atmosphere generally agree well (for example, the model spread in the TOA flux is below \SI{7}{\percent}).
Radiative heating rates in the atmosphere are also rather similar between the models in their respective cases; the models disagree substantially only at the very top of the atmosphere.
The tropospheric circulation in all models except for ROCKE-3D is in a Rhines-rotator regime with a strong equatorial superrotation.
The overall strength of the circulation is comparable between the models and the two cases, with the zonal mean zonal winds reaching \SIrange{50}{70}{\m\per\s} and the day-night overturning streamfunction $\Psi^{TL}_{max}$ staying within \SIrange{4.2}{7.4e11}{\kg\per\s}.
Thus, the agreement between the models is relatively better in the dry thermodynamics and radiation, as well as in the atmospheric circulation than it is for moist physics.
While there are certainly important inter-model differences in radiation and dynamics, as we discuss in more detail in Part I of the THAI trilogy \citep{Turbet21_THAI}, we find a much larger discrepancy in the moist aspects of the simulated Hab~1 and Hab~2 climates.

Our experiments show that the largest inter-model differences appear in the water vapor content, cloud amount and distribution.
The wettest troposphere is simulated by ExoCAM both in the Hab~1 and Hab~2 cases, which can be attributed to its overall warmer climate.
In the stratosphere, the clear outlier is LMD-G, predicting specific humidity at least an order of magnitude lower than that in the other three models (VMR\textsubscript{\SI{1}{\hecto\pascal}} in Table 2).
This outcome contradicts previous studies, in which ExoCAM and LMD-G were used to simulate fast-rotating planets orbiting a G-dwarf star \citep{Wolf2015} and tidally locked planets orbiting an M-dwarf \citep{Yang2019}.
As mentioned in Sec.~\ref{sec:hab1_thermo}, the main causes of LMD-G's dry stratosphere are the cold tropopause and the inability of the convective adjustment scheme to account for moisture detrainment beyond the level of neutral buoyancy at the tropopause.
Additional evidence for the latter is the scant amount of clouds at high altitudes (in a shape of an ``anvil'') above the substellar point (see Sec.~\ref{sec:hab1_clouds}).
As a result, the water absorption band at \SI{\approx 6}{\micro\m} is much weaker in LMD-G than that in the simulations by other models \citep[see Paper III:][]{Fauchez21_THAI}.

Not only does LMD-G have the driest upper atmosphere, its cloud amount is consistently the smallest among THAI models.
This contradicts the Earth-like simulations of \citet{Wolf2015}, but generally supports the more recent simulations of \citet{Yang2019}, when compared to ExoCAM (then called CAM4\_Wolf).
The cloudiest simulation, in terms of the global mean cloud fraction and integrated cloud condensate, is produced by ROCKE-3D.
Similar to the UM simulations of TRAPPIST-1e in \citet{Sergeev2020}, increased cloud fraction as well as the cloud condensate concentrated at the substellar point, are congruent with a change in the tropospheric circulation regime.
Unlike the other three models, ROCKE-3D favors the accumulation of eastward momentum in two zonal jets in the extratropics instead of the equator.
Such a regime of the two extratropical jets was shown by \citet{Turbet:2018aa} for a case of an \ce{N2} atmosphere with \SI{10}{\percent} \ce{CH4} in their 3D simulations of TRAPPIST-1e, f, h, and g.
This suggests that a change in the gaseous absorption and a resulting change in the static stability of the atmosphere may induce a change in circulation, which, in turn, modulates the cloud cover.

Cloud metrics shown in this paper also demonstrate that one must be cautious when interpreting GCM output, as has been also noted by \citet{Kopparapu2017}.
Cloud area fraction can give a general impression of how ``cloudy'' a planet is and thus how reflective its atmosphere, i.e. the albedo.
At the same time, this parameter is highly model-dependent.
Moreover, it is usually insensitive to the optical thickness or the altitude of the cloud and can give a false impression of its radiative importance.
One solution is to filter optically thin clouds \citep{Boutle2017}.
Another is to show vertically integrated cloud content, i.e. the cloud water path \citep[e.g.][]{Kopparapu2017,Wolf2017,Wolf2019}.
The difference between these two diagnostics can be seen in Figs.~\ref{fig:hab1_cloud_maps} and \ref{fig:hab2_cloud_maps}.
For example, in the ROCKE-3D simulation of the Hab~1 case almost all grid boxes have a cloud fraction \SI{>50}{\percent} (Fig.~\ref{fig:hab1_cloud_maps}g), but the cloud water path is quite low in the tropical band on the night side of the planet (Fig.~\ref{fig:hab1_cloud_maps}c), indicating that that region is covered by optically thin high clouds (Fig.~\ref{fig:hab1_cloud_cross}c,g,k).
This demonstrates how looking at only one cloud diagnostic in a GCM can be misleading and should be taken into account when connecting GCM output to atmospheric retrieval tools.

Clouds are a serious obstacle to atmospheric characterization, because they raise the altitude of the continuum level of a transmission spectrum \citep[e.g.][]{Fauchez:2019}, therefore reducing the amplitude of the absorption lines.
Even though ROCKE-3D is cloudier overall than other GCMs in both Hab~1 and Hab~2 cases, its cloud condensate is concentrated at the substellar point (Fig.~\ref{fig:hab1_cloud_maps}c, \ref{fig:hab2_cloud_maps}c).
At the terminators, the largest cloud water path is simulated by ExoCAM.
Furthermore, the cloud mixing ratio maximum in ExoCAM reaches its maximum at the highest altitude among the THAI models, especially at the eastern terminator, where the cloud content is higher.
Consequently, its continuum level is the highest and, compared to other models, requires a higher number of transits to detect \ce{CO2} absorption peaks using JWST \citep[see Paper III,][]{Fauchez21_THAI}.
The UM, on the other hand, tends to produce lower cloud decks at the terminators with smaller temporal variability than that in the ExoCAM, which means it has the lowest continuum level and requires the smallest number of orbits for \ce{CO2} detection.
Accurate representation of cloud opacity and cloud top heights, including their spatial variability, has a substantial impact on the detectability estimates, making 3D GCMs a more appropriate tool than 1D models with a fixed cloud layer \citep[e.g.][]{Lin21}.
The inter-model differences in cloudiness of the terminator stem from the differences in how GCMs distinguish between convective and stratiform clouds.
For example, if one GCM produces a lot of convective clouds, which would be concentrated at the substellar point, and few stratiform clouds, the terminator region in its simulation would be less cloudy.
While this separation between cloud types is artificial, it is common for terrestrial GCMs and thus should be investigated further in future intercomparisons.

A key potentially observable feature of cloudy dynamic atmospheres is the time variability of opacity at the terminators.
This variability will most likely stay below the noise level of JWST's NIRSpec PRISM instrument and not affect retrieved abundances \citep{May2021water}.
While our simulations support \citet{May2021water}'s finding, they provide a multi-model estimate and may prove useful if a similar planet is found closer to Earth or if TRAPPIST-1e is probed by better instruments \citep{Fauchez21_THAI}.
As discussed in Sec.~\ref{sec:hab1_time_var} and \ref{sec:hab2_time_var}, ExoCAM has the highest amplitude of cloud variability among the THAI models, and this corresponds to the highest variability in the atmospheric transit depth relative to other GCMs, as shown for different wavelengths of the transmission spectrum in Paper III \citep{Fauchez21_THAI}.
The periodicity of ExoCAM's cloud variability is approximately 12.5 orbits, which is comparable to the minimum number of transits required to detect \ce{CO2} features and thus may need to be taken into account if TRAPPIST-1e is observed during successive transits.
Differences between the simulations of the Hab~1 and Hab~2 climate lead to an increased amplitude of the cloud variability, which affects ExoCAM and the UM the most \citep{Fauchez21_THAI}.
Note, however, that for other models the differences between the variability in the spectra between the Hab~1 and Hab~2 cases are muted, likely because the highest clouds have a bigger impact than the total cloud condensate in the column. 

\subsection{Habitability and climate stability} \label{sec:habitability}
Differences in the mean climate shown in Sec.~\ref{sec:results} imply that the prediction of habitability of TRAPPIST-1e is model-dependent.
With respect to the inner edge of the habitable zone, none of the THAI simulations enter the water loss stage or a runaway greenhouse state.
Stratospheric water content in the Hab~1 case, taken here as the \ce{H2O} volume mixing ratio at the \SI{1}{\hecto\pascal}, is about \SIrange{5}{50e-8}{\mol\per\mol} (Table~\ref{tab:glob_diag}) and stays comfortably below the classical water loss limit of \num{3e-3} \citep{Kasting93}, especially in the case of LMD-G.
The inter-model difference in the stratospheric water content is an order of magnitude (due to LMD-G having a much lower value), which for a warmer climate, e.g. a planet orbiting the star closely such as TRAPPIST-1d, may result in the eventual loss of water on geological time scales.
While the atmosphere is substantially warmer and thus moister in the Hab~2 simulations relative to Hab~1, the stratospheric volume mixing ratio for the Hab~2 case is still 1--2 orders of magnitude lower than the limit --- similar to the fast-rotating simulations for a \SI{2600}{\K} star demonstrated by \citet{Kopparapu2017}.
Note, however, that the \citet{Kasting93} limit is a conservative estimate derived from a cloudless 1D Earth simulation.

The Hab~1 and Hab~2 simulations also avoid the other extreme, namely the \ce{CO2} condensation limit, which can trigger a runaway atmospheric collapse \citep{Turbet:2018aa}.
In the coldest simulation --- the UM's simulation of the Hab~1 case --- the night side exhibits a strong temperature inversion in the boundary layer, exceeding \SI{30}{\K} in the lowest \SI{100}{\m}.
Even in the coldest places on Earth, the inversion rarely reaches this strength, counteracted by turbulent mixing by topography-induced mesoscale eddies \citep{Joshi20}.
Our simulations have flat topography and therefore potentially overestimate the surface cooling in the night-side cold traps.
Even with this overestimation, the lowest temperature stays above the condensation point of \ce{CO2}, both for the Hab~1 (\SI{\approx 125}{\K}) and in Hab~2 cases (\SI{\approx 195}{\K}).
Note that the Hab~1 surface temperature minimum in the UM is 20--30~K lower than that in the other three GCMs (see Figs.~\ref{fig:hab1_t_sfc_map}d and \ref{fig:hist_t_sfc}a), so in a colder state it would be closer to a potential atmospheric collapse, all other factors being equal.

Within these extremes demarcating the habitable zone boundaries, habitability of the temperate climate of TRAPPIST-1e can be defined by the abundance of stable surface liquid water.
We can use the planet-average open ocean fraction as a metric of this abundance, with the obvious caveat of the lack of a dynamic ocean in our simulations \citep{DelGenio2019}.
Fig.~\ref{fig:hab1_tseries}d demonstrates a fairly close estimate of the ice-free ocean area at around 0.20--0.23 in the GCMs.
Despite the fact that the UM predicts a lower surface temperature both globally and on the day side, its open ocean fraction is essentially the same as that in ExoCAM (0.23).
LMD-G and ROCKE-3D have a similar day-side average surface temperature, and their ocean fraction fluctuates around $0.20$.

While being relatively close to entering the water loss limit, Hab~2 simulations predict a larger fraction of deglaciated habitable area at the surface: the surface is warmer and the day-night contrast is roughly twice as small as in Hab~1.
The ice-free fraction varies from 0.36 in LMD-G to $\approx0.75$ in the UM and ROCKE-3D to 1 in ExoCAM (Fig.~\ref{fig:hab2_tseries}d).
Thus, in line with a larger inter-model spread in the global mean surface temperature, the ice-free fraction disagreement in the Hab~2 case is larger than that in Hab~1.
This is a result of the surface temperature distribution with respect to the freezing point of water being different in the two experiments (Fig.~\ref{fig:hist_t_sfc}).
In the Hab~1 simulations, the freezing point is closer to the right tail of this distribution, where the agreement between the models is higher; for Hab~2 it is closer to the left tail of the distribution (i.e. the night side temperatures), where models have the largest discrepancy.

\subsection{Caveats and future work}
We have used four state-of-the-art 3D GCMs to re-examine the climate of TRAPPIST-1e in a more structured way than previous studies.
Nevertheless, our study has some caveats.
First, we chose a simplified gas composition for both the Hab~1 and Hab~2 cases: \ce{N2}- and \ce{CO2}-dominated atmosphere with \ce{H2O} as the condensible species.
Furthermore, we chose a total atmospheric pressure of 1 bar.
While these setups cover two key types of atmospheric composition, not least for the sake of easier comparison with the climate dynamics on Earth, TRAPPIST-1e may possess a completely different atmosphere \citep{Turbet20b} and more 3D GCM experiments are needed to widen the parameter space \citep[e.g.][]{Fauchez:2019,ZhangYang20}.

Second, the Hab~1 and Hab~2 experiments presented here are predicated on the assumption of abundant water at the surface of the planet.
It is possible that TRAPPIST-1e has lost all water throughout its lifetime, but some studies suggest that the hydrodynamic escape may have stopped once it entered the habitable zone \citep{Bolmont17,Bourrier17}.
Additionally, late-stage outgassing of the mantle may have replenished water on the surface \citep{Moore20}.
TRAPPIST-1e may never enter a water-loss phase and retain its oceans over a timescale of 100 Gyr \citep[][]{Bolmont17,Kopparapu2017}.
Thus, together with the dry planet scenarios explored in Paper I \citep{Turbet21_THAI}, scenarios with abundant water shown in the present paper cover the two main pathways for rocky planets orbiting M-dwarfs \citep{Tian:2015}.

Third, the surface is assumed to be completely covered with a static slab ocean without any continents.
The presence of continents and orography may significantly change the planet's hydrological cycle \citep[e.g.][]{Lewis_2018} and boundary layer mixing on the night side \citep{Joshi20}.
In some cases and in the absence of large continents, the oceanic heat transport may also change the simulated climate \citep[e.g.][]{DelGenio2019}, but the oceanic component of exoplanet models requires a separate intercomparison \citep{Fauchez2021_THAI_workshop}.

Fourth, it is unknown which model is closer to the truth, since there are no observations of the atmosphere of TRAPPIST-1e available yet.
However, the present study and similar future multi-model intercomparisons helps to identify biases of one GCM relative to others, whether in a specific scenario or systematically across the parameter space.
Before observations of temperate rocky exoplanets become available, high-resolution cloud-resolving models or large eddy simulations could be used to validate moist physics parameterizations of exoplanet GCMs \citep[e.g.][]{Sergeev2020,Lefevre21}.
More detailed intercomparisons of individual parameterizations of convection, boundary layer, and microphysics are also needed.

Future studies of temperate climates from a 3D perspective are encouraged to build upon the experiments analyzed in this paper.
It is pertinent to address the obvious caveats mentioned above, as well as expand this effort both in the parameter space and by comparing more GCMs.
More habitable scenarios should be explored by varying the atmospheric composition and pressure of TRAPPIST-1e.
This has been started in a few previous works \citep[e.g.][]{Wolf2017,Turbet:2018aa,Fauchez:2019}, but requires a coordinated multi-model approach, several benefits of which have been demonstrated by the present study.
This type of study would bring best value if performed periodically (e.g. every 5 years), similar to the CMIP framework \citep[e.g.][]{Eyring16}.
We note, however, that future intercomparisons of exoplanet GCMs should find a balance between increasing the model complexity and reducing the inter-model spread.

\section{Conclusions} \label{sec:conclusions}
In this Part II of the THAI study, we analyzed a moist climate of TRAPPIST-1e in an aquaplanet configuration simulated by the four 3D GCMs --- ExoCAM, LMD-G, ROCKE-3D and the UM.
We hope this study will propel future model intercomparisons of potentially habitable extrasolar worlds.
Our conclusions are as follows.
\begin{enumerate}
    \item%
Despite many inter-model differences, Hab~1 and Hab~2 climates in all four models stay within the habitability limits.
They sustain an active hydrological cycle without entering the water loss stage: the stratospheric water content is at most $\mathcal{O}(\SI{e-7}{\mol\per\mol})$ in the Hab~1 case and $\mathcal{O}(\SI{e-5}{\mol\per\mol})$ in the Hab~2 case --- below the classical water loss limit.
The lowest surface temperature is also above the \ce{CO2} condensation limit and thus the condition for night-side atmospheric collapse is not triggered.
    \item%
The global mean surface temperature is \SIrange{232}{246}{\K} in the Hab~1 simulations and \SIrange{271}{295}{\K} in the Hab~2 simulations, with a relatively larger inter-model spread on the night side than on the day side.
The models agree more in their estimate of the ice-free area fraction for Hab~1 (0.20--0.23) and less for Hab~2 (0.36--1.00).
    \item%
The largest inter-model differences exist in the variables directly affected by moist physics parameterizations.
LMD-G produces the driest stratosphere, especially in the Hab~1 case due to a ``cold trap'' at the tropopause.
ExoCAM has the wettest troposphere, due to it being substantially warmer, which is a consequence of strong stellar flux absorption by \ce{CO2}.
Globally averaged, LMD-G is also consistently less cloudy, both in terms of mean cloud area fraction and column-integrated cloud content.
Note that this is not the case at the terminators \citep[for further discussion, see Part III][]{Fauchez21_THAI}.
The cloudiest Hab~1 and Hab~2 climates are predicted by ROCKE-3D, which simulates TRAPPIST-1e enveloped by clouds with a high concentration of cloud water at the substellar point.
    \item%
Tropospheric circulation in three models (ExoCAM, LMD-G and the UM) is characterized by a Rhines-rotator regime with a strong equatorial superrotation both in Hab~1 and Hab~2 cases.
ROCKE-3D is an outlier and predicts a zonal circulation dominated by two extratropical jets.
A different circulation regime in ROCKE-3D partly explains the differences in cloud patterns relative to other GCMs.
    \item%
The models exhibit different amplitude and periodicity of atmospheric variability.
At the terminators, ExoCAM has a distinctly higher amplitude of cloud variability than other models, and a period of 12.5 orbits.
In other models, the variability is generally of higher frequency ($\approx$2.5--11 orbits).
The amplitude grows in Hab~2 relative to Hab~1, especially in the ExoCAM and UM cases.
\end{enumerate}

\acknowledgments
The authors are grateful to the two anonymous reviewers whose comments helped to improve this paper.
D.E.S., I.A.B., F.H.L, J.M. and N.J.M. acknowledge use of the Monsoon2 system, a collaborative facility supplied under the Joint Weather and Climate Research Programme, a strategic partnership between the Met Office and the Natural Environment Research Council.
We acknowledge support of the Met Office Academic Partnership secondment program.
This work was partly supported by a Science and Technology Facilities Council Consolidated Grant (ST/R000395/1), UKRI Future Leaders Fellowship (MR/T040866/1), and the Leverhulme Trust (RPG-2020-82).
T.F. and M.J.W. acknowledge support from the GSFC Sellers Exoplanet Environments Collaboration (SEEC), which is funded in part by the NASA Planetary Science Divisions Internal Scientist Funding Model.
This project has received funding from the European Union's Horizon 2020 research and innovation program under the Marie Sklodowska-Curie Grant Agreement No. 832738/ESCAPE.
M.T. thanks the Gruber Foundation for its generous support to this research.
M.T. was granted access to the High-Performance Computing (HPC) resources of Centre Informatique National de l'Enseignement Sup\'erieur (CINES) under the allocations \textnumero~A0020101167 and A0040110391 made by Grand \'Equipement National de Calcul Intensif (GENCI).
This work has been carried out within the framework of the National Centre of Competence in Research PlanetS supported by the Swiss National Science Foundation.
M.T. acknowledges the financial support of the SNSF.
M.T. acknowledges support from the PORTAL BRAIN-be 2.0 BELSPO project.
M.T. and F.F. thank the LMD Generic Global Climate team for the teamwork development and improvement of the model.
M.J.W was supported by NASA's Nexus for Exoplanet System Science (NExSS). Resources supporting the ROCKE-3D simulations were provided by the NASA High-End Computing (HEC) Program through the NASA Center for Climate Simulation (NCCS) at Goddard Space Flight Center.
J.H.M. acknowledges funding from the NASA Habitable Worlds program under award 80NSSC20K0230.

The authors acknowledge the help of Andrew Ackerman to set up the cloud diagnostics in ROCKE-3D.

The THAI GCM intercomparison team is grateful to the Anong's Thai Cuisine restaurant in Laramie for hosting its first meeting on November 15, 2017.

Numerical experiments performed for this study required the use of supercomputers, which are energy intensive facilities and thus have non-negligible greenhouse gas emissions associated with them.
We estimate that the final versions of Hab~1 \& Hab~2 experiments resulted in roughly \SI{\approx 1280}{\kg} \ce{kgCO2e}, including \SI{\approx 520}{\kg} from ExoCAM runs, \SI{\approx 80}{\kg} from LMD-G, \SI{\approx 650}{\kg} from ROCKE-3D and \SI{\approx 30}{\kg} from the UM.
These numbers are the product of the following three factors: i) CPU-hours used by each of the model to reach the steady state, ii) kWh per CPU-hour, which depends on the efficiency of the supercomputer used, and iii) \ce{kgCO2e} per kWh of energy, which varies a lot from country to country and so is the biggest source of uncertainty in our estimates.

\software{
Scripts to process and visualize THAI data are available on GitHub: \url{https://github.com/projectcuisines/thai_trilogy_code} and are dependent on the following Python libraries: \texttt{aeolus} \citep{aeolus}, \texttt{iris} \citep{iris}, \texttt{jupyter} \citep{jupyter}, \texttt{matplotlib} \citep{Hunter2007}, \texttt{numpy} \citep{numpy}, \texttt{windspharm} \citep{windspharm}, \texttt{xarray} \citep{xarray}.
ExoCAM \citep{Wolf2015} is available on GitHub: \url{https://github.com/storyofthewolf/ExoCAM}.
LMD-G is available upon request from Martin Turbet (martin.turbet@lmd.jussieu.fr) and François Forget (francois.forget@lmd.jussieu.fr).
ROCKE-3D is public domain software and available for download from \url{https://simplex.giss.nasa.gov/gcm/ROCKE-3D/}.
Annual tutorials for new users take place annually, whose recordings are freely available on line at \url{https://www.youtube.com/user/NASAGISStv/playlists?view=50&sort=dd&shelf_id=15}.
The Met Office Unified Model is available for use under licence; see \url{http://www.metoffice.gov.uk/research/modelling-systems/unified-model}.
          }

\section{Data accessibility}
All our GCM THAI data are permanently available for download here: \url{https://ckan.emac.gsfc.nasa.gov/organization/thai}, with variables described for each dataset. If you use those data please cite the current paper and add the following statement: ``THAI data have been obtained from \url{https://ckan.emac.gsfc.nasa.gov/organization/thai}, a data repository of the Sellers Exoplanet Environments Collaboration (SEEC), which is funded in part by the NASA Planetary Science Divisions Internal Scientist Funding Model.''\\

\bibliography{biblio}{}
\bibliographystyle{aasjournal}



\end{document}